\documentclass[aps,prb,footinbib,showpacs,showkeys,amssymb,amsmath,superscriptaddress,twocolumn]{revtex4-1}
\pdfoutput=1
\usepackage{graphicx,graphics}
\usepackage[export]{adjustbox}
\usepackage{verbatim}
\usepackage{braket}
\usepackage{float}
\usepackage{breqn}
\usepackage{subfig}
\usepackage{color}
\usepackage{hyperref}
\usepackage{enumerate}

\makeatletter
\let\cat@comma@active\@empty
\makeatother

\begin{document}
%	\title{Criticality in non-Hermitian Su-Schrieffer-Heeger model with long-range couplings}
	\title{Multi-criticality and long-range effects in non-Hermitian topological models}
	
	\author{Y R Kartik}
		\thanks{yrkartik@gmail.com}
	\affiliation{Theoretical Sciences Division, Poornaprajna Institute of Scientific Research, Bidalur, Bengaluru 562164, India.}
	\affiliation{Graduate Studies, Manipal Academy of 
	Higher Education, Madhava Nagar, Manipal-576104, India.}
	\author{Sujit Sarkar}
		\thanks{sujit.tifr@gmail.com}
	\affiliation{Theoretical Sciences Division, Poornaprajna Institute of Scientific Research, Bidalur, Bengaluru 562164, India.}
	 
	\date{\today} 
	
\begin{abstract}
	Long-range effects induce some interesting behavior and considered as a gateway to understand the non-local behavior in the quantum systems. Especially, the long-range topological models became a platform for the realization of new quasi-particles, which are believed to be potential candidates for the topological qubits. In this work, we consider non-Hermitian Su-Schriffer-Heeger (SSH) model and discuss the interplay of non-Hermiticity and long-range effects. We use the approach of momentum space characterization, critical exponents and curvature renormalization group (CRG) method to understand the aspects of interplay. The longer-range (finite neighbors) effect produces higher winding numbers, where we observe a staircase of transitions among the even-even and odd-odd winding numbers which depends on the number of interacting neighbors. Here we also highlight the effect of multi-criticality in the system and show that they belong to a different universality class. The interplay of long-range (infinite neighbors) effect and non-Hermiticity produces fractional topological invariants, and we analyze them from the behavior of pseudo-spin vectors. We also determine the long-range and short-range limit of the model through universality class of critical exponents. Our work mainly showcases that how the study of criticality in topological system is interesting in exploring the interplay of non-Hermiticity and long-range effects.
%With a motivation to understand the interplay of non-Hermiticity and long-range effects in topological system, we conduct a theoretic study of chiral Su-Schriffer-Heeger (SSH) chain with imbalanced intracell hopping. We use the approach of momentum space characterization, universality class of critical exponents (UCCE) and curvature renormalization group method (CRG) to explore the criticality. We find some interesting behavior in the interface of long-range short-range limits where the UCCE is different than the rest of the parameter space. To understand the effect of neighboring couplings on the topological system, we consider longer-range (finite neighbors) and long-range (infinite neighbors) chains where we observe staircase of topological transitions and polylogarithmic nature respectively. The CRG approach effectively recognizes the fixed, critical point configuration, where the UCCE are distinct from the rest of the parameter space.  Our work mainly showcases that how the study of criticality in topological system is interesting in exploring the interplay of non-Hermiticity and long-range effect.
\end{abstract}
	
\maketitle
\section{Introduction}
Topological state of matter is the milestone in the area of condensed matter, both from the perspective of theory and experiments\cite{hasan2010colloquium,moore2010birth,stanescu2016introduction,bernevig2013topological,ortmann2015topological,haldane2017nobel}. The transitions among the topological phases can not be apprehended by the conventional Landau theory of symmetry breaking, as they do not possess an order parameter\cite{landau1965collected,berry1985classical,wilczek1989geometric,wen2016introduction}. The topological phases possess localized edge modes which are protected by certain symmetries and are robust against the external perturbations\cite{niu2012majorana,sarma2015majorana,kitaev2001unpaired}. Topological invariant is the most accepted tool to differentiate the index of the topological phases, where topological invariant is equal to the number of localized edge modes\cite{alecce2017extended}. Even though topological invariant can not be considered as local order parameter, it has a well defined form in gapped phases and ill-defined at criticality. The edge mode contains a topological localization length, which diverges as the system drives towards criticality\cite{continentino2020finite,chen2019topological,molignini2020unifying,molignini2018universal,chen2016scaling,chen2017correlation,malard2020scaling,chen2016scaling2,chen2018weakly,chen2019universality,molignini2020generating,abdulla2020curvature,molignini2020unifying,kumar2022topological}. This diverging nature gives the idea of explaining the criticality through set of critical exponents and renormalization group methods\cite{kempkes2016universalities,rufo2019multicritical,kartik2021topological,kumar2020multi}. Especially renormalization group (RG) method can be very helpful during the topological transitions where there are actually no order parameters\cite{sachdev2007quantum}.\\
Introducing non-Hermiticity in topological systems opened a new doorway towards the unexplored parts of novel phases of matter \cite{ashida2020non}. The works in non-Hermitian topological systems from the perspective of bulk-boundary correspondence\cite{jin2019bulk}, loss-gain\cite{agarwal2021pt}, dissipation\cite{rahs}, symmetry\cite{shiozaki2021symmetry,wang2015spontaneous,kawabata2019symmetry,kawabata2018parity}, skin effect\cite{yao2018edge} and critical scaling\cite{arouca2020unconventional} gained a lot of attention in recent days. The realization of non-Hermitian models in experimental setups provided new understandings in optical systems\cite{schomerus2010quantum,el2019dawn}, electric circuits\cite{hofmann2020reciprocal,zhang2020non} and ultra-cold atomic gases\cite{jotzu2014experimental,bloch2012quantum}. The longer-range effects in non-Hermitian topological system created some fractional topological phases including non metallic phases\cite{bao2021exploration,yin2018geometrical}, which gained attention both theoretically and experimentally \cite{xu2021coexistence,li2018extended,bao2021exploration}.\\
%This development attracted the other areas like quantum information\cite{amin2019information,stanescu2016introduction}, spintronics\cite{bibid}, trapped ion\cite{bibid}, ultracold atoms\cite{bibid} with great interest. However, the criticality and bulk-boundary correspondence still remain as the hot topic under different conditions\cite{kopp2005criticality,rahul2021majorana}. There are significant works to explain criticality from critical exponents\cite{continentino2017quantum,continentino2020finite,rufo2019multicritical,chen2008intrinsic,chen2019topological}, renormalization group\cite{kumar2020multi}, entanglement entropy and central charge\cite{kumar2021topological} for both Hermitian and non-Hermitian topological systems.\\
%Non-Hermitian effects in quantum systems are the area of curiosity due to their significant role in understanding the physics of gain-loss\cite{rahs}, skin effect\cite{yao2018edge}, spectral degeneracies\cite{bibid} and problems of open quantum systems\cite{bibid}. Introduction of non-Hermiticity in topological state of matter created a big doorway to understand the non-Hermitian novel phases of matter which can not be apprehended by the Landau theory of symmetry breaking\cite{landau1965collected,stanescu2016introduction}. The efforts of criticality, field theory, bulk-boundary correspondence, non-equilibrium dynamics and quenching studies in non-Hermiticity made the field more stable and versatile\cite{kawabata2018parity,kawabata2019symmetry}. \\
The longer-range models are important milestones in topological systems due to the formation of more than one edge modes and formation of multi-critical points\cite{kartik2021topological}. The increase in the number of coupling neighbor generates higher winding numbers (WNs). Furthermore, the increase in the decay parameter reduces the system towards short-range through staircase of topological transition\cite{kartik2021topological}. Multi-criticality is an interesting phenomenon which can exhibit a combined behavior of two or more different criticalities\cite{kumar2020multi,kumar2021topological,kartik2021topological}. In Hermitian systems, multi-critical point witnesses transition between different (gapped-gapped, gapless-gapless, gapped-gapless) phases\cite{kartik2021topological}. In addition, under some cases, multi-critical points have been recognized as the point where Lorentz invariance violates\cite{kumar2020multi}. Also, there are observations, where the introduction of non-Hermiticity violates multi-critical effects\cite{rahs}.\\
On the other hand, long-range effects in condensed matter has a long history and had been an area of curiosity over the decades\cite{defenu2021long}. Especially, long-range effects in topological systems been used to understand the non-local behavior of the fermions\cite{vodola2015correlations}. 
Long-range topological models exhibit the emergence of massive edge modes\cite{vodola2014kitaev}, topological transition without gap closing\cite{vodola2014kitaev,kartik2021topological}, area law violation for Von Neumann entropy and breaking of Lorentz invariance\cite{vodola2015long}. The long-range models are also important as they can be realized in the experiment setups of trapped ions\cite{deng2005effective,britton2012engineered,hauke2010complete,roy2019quantum}, multi-mode cavities\cite{douglas2015quantum}, magnetic impurities\cite{zhang2019majorana,menard2015long} and simulated circuits\cite{amin2019information}. Even in two dimensional models, long-range effects has enhanced the topological behaviors\cite{viyuela2018chiral}. The massive edge modes has been predicted to be a potential qubit for topological computation\cite{viyuela2016topological}. The extension of long-range effects towards the non-Hermitian systems naturally triggers the curiosity to explore the interplay among them.
Here our motivation is as follows.
\begin{itemize}
\item To understand the behavior of non-Hermitian system with increased neighboring coupling.
\item  To explore the multi-critical effects in non-Hermitian systems and to differentiate the multi-criticality from normal criticalities through universality class and curvature renormalization group (CRG).
\item To understand the interplay of non-Hermiticity and long-range effect in a topological system, thus to understand the non-local behaviors of the fermions in these systems. 
\item The interface of short-range and long-range limit is an interesting topic of discussion. There are many arguments to decide the short-range limit based on entanglement entropy\cite{vodola2014kitaev}, conformal invariance\cite{vodola2015long} and bulk-edge correspondence\cite{lepori2017long}. Here we give an approach of universality class of critical exponents to decide the short-range limit in a non-Hermitian long-range system.
\end{itemize}
With these motivations, we present the manuscript in following pattern. In Sec.\ref{sec2} we introduce the model Hamiltonian and explain the momentum space properties. This section includes the study of universality class of critical exponents, CRG method and behavior of multi-critical points in longer-range model. In Sec.\ref{sec3}, we study on long-range models with winding vector analysis, universality class, higher order quantum transitions and interplay among non-Hermiticity and long-range effect. In Sec.\ref{sec4} we give outlook and experimental possibility of our work and conclude in Sec.\ref{sec5}.
\section{Model Hamiltonian and properties}\label{sec2}
We consider chiral non-Hermitian 1D Su-Schriffer-Heeger (SSH) chain with long-range interaction. Here, the non-Hermiticity is due to the imbalanced intracell hoppings, i.e., the hopping within the lattice $i$ (between the sub-lattice $A_i,B_i$) is a real non-reciprocal quantity, which violates the Hermiticity in the Hamiltonian\cite{yin2018geometrical}.
\begin{eqnarray}
	H&=&\sum_{j=1}^{L-l}
	(t-\delta)c^{\dagger}_{j,a}c_{j,b}+(t+\delta)c^{\dagger}_{j,b}c_{j,a}\nonumber\\
	&+&\sum_{j=1}^{L-l}
	\sum_{l=1}^{r} \frac{t^{\prime}}{l^{\alpha}}(c^{\dagger}_{j+l,a}c_{j,b}+c^{\dagger}_{j,b}c_{j+l,a}),\label{Ham}	
\end{eqnarray}
where $\frac{t^{\prime}}{l^{\alpha}}$ is the intercell long-range hopping with $l$ as site index and $\alpha$ as decay parameter respectively. The imbalance in the intracell coupling is introduced through the non-zero term $\delta$, which breaks the Hermiticity and leads to non-Hermitian skin effect\cite{yao2018edge,yin2018geometrical}. The term $L$ is the system size where the intercell coupling occurs between sites $i$ and $i+l$. The term $l$ can take the upper limit $r$, where $r$ can be infinite. In such cases, it is called long-range systems with infinite number of coupling neighbors. On the other hand, if $r$ is finite, then it is longer-range with finite number of coupling neighbors. In addition to these factors, as $\alpha\rightarrow\infty$, the model reduces to short-range version irrespective of number of interacting neighbors.
\begin{figure}[H]
	\centering
	\includegraphics[width=\columnwidth,height=4cm]{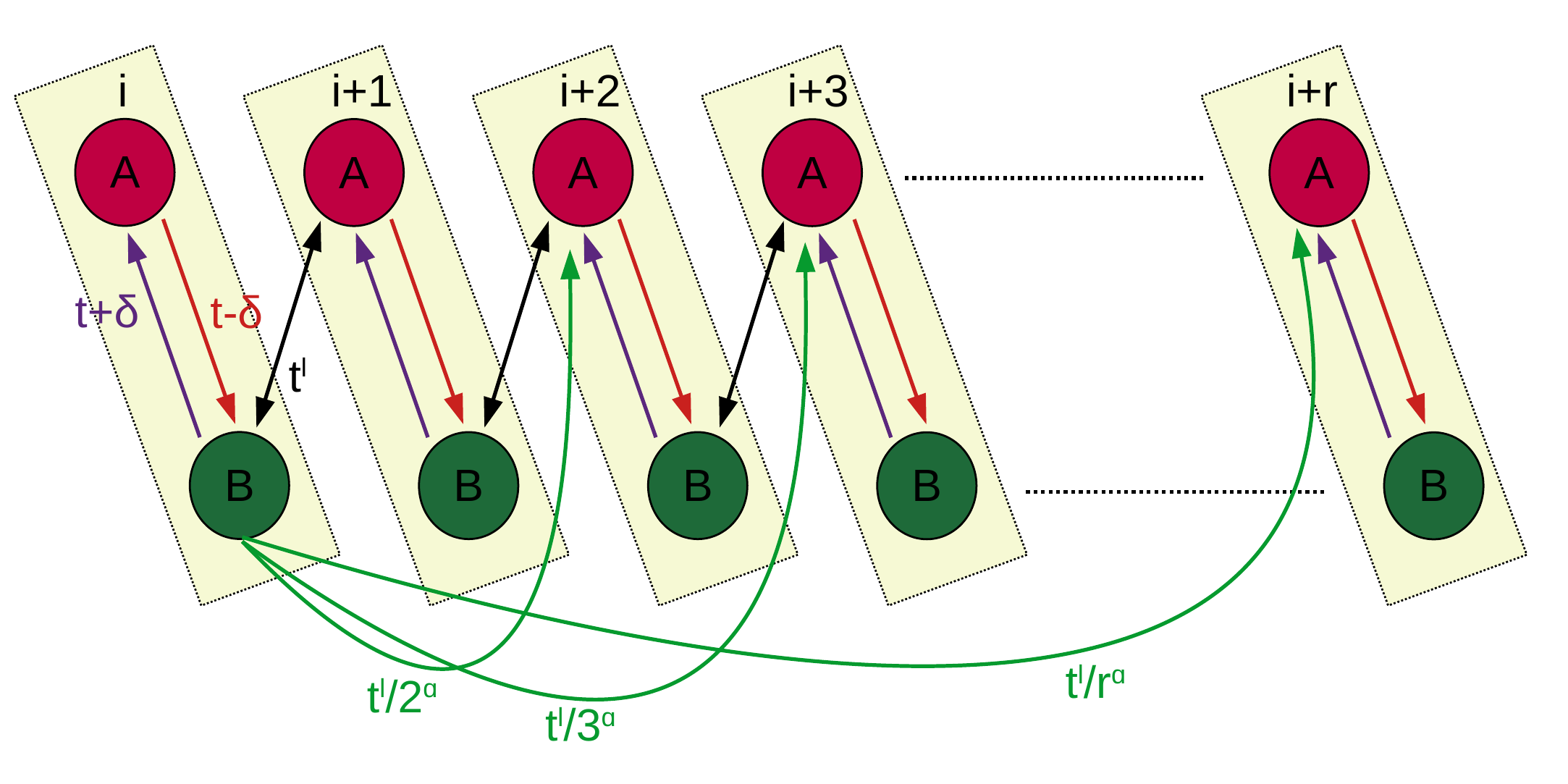}
	\caption{(Color online) Schematic representation of non-Hermitian long-range SSH chain with open boundary condition. Colored rectangles represent the unit cell, $t+\delta$ and $t-\delta$ are the forward and backward hopping (intracell hopping) between sub-lattice sites $A$ and $B$ respectively. Here $t^{\prime}$ is the intercell hopping (short-range limit) and $\frac{t^{\prime}}{r^{\alpha}}$ is the longer-range hopping parameter.}
	\label{fig1}
\end{figure}
After the Fourier transformation, BdG Hamiltonian can be written as 
\begin{equation}
	H_{BdG}(k)=\left(\begin{matrix}
		0&& \chi_x(k)+i\chi_y(k)\\
		\chi_x(k)-i\chi_y(k)&& 0\\
	\end{matrix}
	\right).\end{equation}\label{matrix}	
With the use of Anderson 
pseudo-spin approach~\cite{anderson1958coherent,sarkar2018quantization}, 
we write the BdG Hamiltonian in the pseudo-spin basis as 
\begin{dmath}
	H_{BdG}(k)=\chi_x(k) \vec{\sigma_x}+\chi_y(k)\vec{\sigma_y}, 
\end{dmath}\label{pseudo}
where $\sigma_x,\sigma_y$ 
are the Pauli matrices and the corresponding coefficients,
\begin{eqnarray}
	\chi_x(k)&=& t +t^{\prime}\sum_{l=1}^{r}\frac{\cos
		(kl)}{l^{\alpha}},\nonumber\\
	\chi_y(k)&=&t^{\prime}\sum_{l=1}^{r}\frac{\sin(kl)}{l^{\alpha}}-i\delta.
	\label{pseudospin}
\end{eqnarray} 
It is to be noted that for $r\rightarrow\infty$ the series involving $\frac{\cos(kl)}{l^{\alpha}}$ and 
$\frac{\sin(kl)}{l^{\beta}}$ terms give rise to polylogarithmic functions \cite{alecce2017extended,vodola2014kitaev}. Energy dispersion is given by\cite{rufo2019multicritical,chen2019topological}
\begin{eqnarray}
	E(k,\mathbf{M})&=&\pm\sqrt{(\chi_z(k))^2+(\chi_y(k))^2},\label{endisp}\\
	&=&\sqrt{(t^{\prime})^2+t^2-\delta^2+2tt^{\prime}\frac{\cos(lk)}{l^{\alpha}}+i2\delta t^{\prime}\frac{\sin(lk)}{l^{\alpha}}}.\nonumber
\end{eqnarray}
Due to the non-Hermiticity, the Eq~\ref{endisp} gives the complex energy spectrum for finite $\delta$\cite{yin2018geometrical}. At the phase boundary the imaginary part vanishes, such that the criticality condition is monitored by the real part. For a non-Hermitian system, the Fermi sea level is contributed by the real part, while the imaginary part signals the decoherence (information leak to the environment)\cite{rahs,yin2018geometrical}.
\subsection{Winding number in non-Hermitian systems}
For topological conditions, the winding vectors form a closed loop in the parameter space, due to the periodicity in the boundary condition. In Hermitian systems, when the parameter $k$ runs from $-\pi$ to $\pi$, which gives topological invariant\cite{wilczek1989geometric},
\begin{equation}
W=\left( \frac{1}{2\pi}\right)\int_{-\pi}^{\pi}\frac{\chi_x\partial_k\chi_y-\chi_y\partial_k\chi_x}{\chi_x^2+\chi_y^2}dk.
\end{equation} 
The imbalance in the intercell hopping induces non-Hermiticity in the Hamiltonian which results in the complex topological angle ($\phi$), i.e., \begin{equation}
	\phi=\phi^{re}+i\phi^{im},
\end{equation}
where the imaginary parts do not contributes to the argument of the topological angle\cite{yin2018geometrical}. But the imaginary part creates two exceptional points (with topological angles $\phi_1,\phi_2$) instead of single Dirac (gap closing) point.
 The angles $\phi_1$ and $\phi_2$ are real and they produce the winding number as\cite{yin2018geometrical}
\begin{equation}
	W=\frac{1}{2}\left( \frac{1}{2\pi}\right) (\oint\partial_k\phi_1dk+\oint\partial_k\phi_2dk),\label{WN}
\end{equation}
where $\phi_1=\tan^{-1}\left( \frac{\chi_y^{re}(k)+\chi_x^{im}(k)}{\chi_x^{re}(k)-\chi_y^{im}(k)}\right) $ and $\phi_2=\tan^{-1}\left( \frac{\chi_y^{re}(k)-\chi_x^{im}(k)}{\chi_x^{re}(k)+\chi_y^{im}(k)}\right) $.(For detailed derivation, refer Appendix~\ref{appa})\\
Here, we initially consider the simplest longer-range model with two neighbor couplings. We gradually increase the number of neighbor couplings to get further longer-range models and to analyze the transitions among them. At the end, we consider infinite couplings to understand the interplay of non-Hermiticity and long-range behavior.
\subsection{Critical exponents and universality class}
\begin{figure}[H]
	\centering
	\includegraphics[width=\columnwidth,height=10cm]{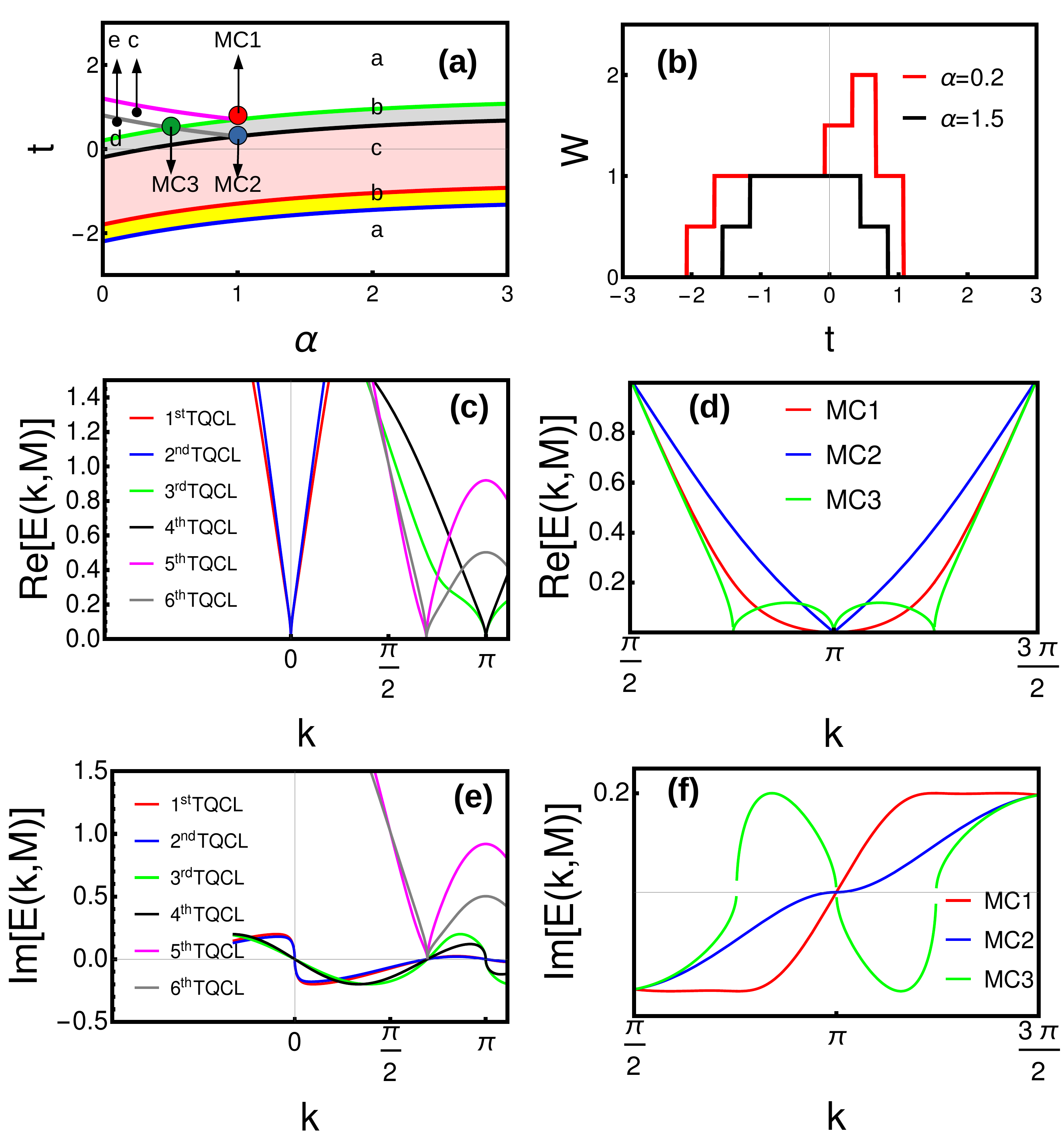}
	\caption{(Color online) (a) Phase diagram for non-Hermitian longer-range SSH chain with two interacting neighbors. Here the red, blue, green, black, magenta and brown lines represent six TQCLs respectively. The points MC1, MC2 and MC3 correspond to first, second and third multi-critical points respectively. The positions $a,b,c,d$ and $e$ represent the topological regions $W=0,1/2,1,3/2$ and $W=2$ respectively. (b) Variation of  winding number against parameter $t$. (c) Energy dispersion (real part) for TQCLs. (d) Energy dispersion (real part) for multi-critical points.(e) Energy dispersion (imaginary part) for TQCLs. (f) Energy dispersion (imaginary part) for multi-critical points.}
	\label{r2}
\end{figure}

\begin{figure}[H]
	\centering
	\includegraphics[width=\columnwidth,height=12cm]{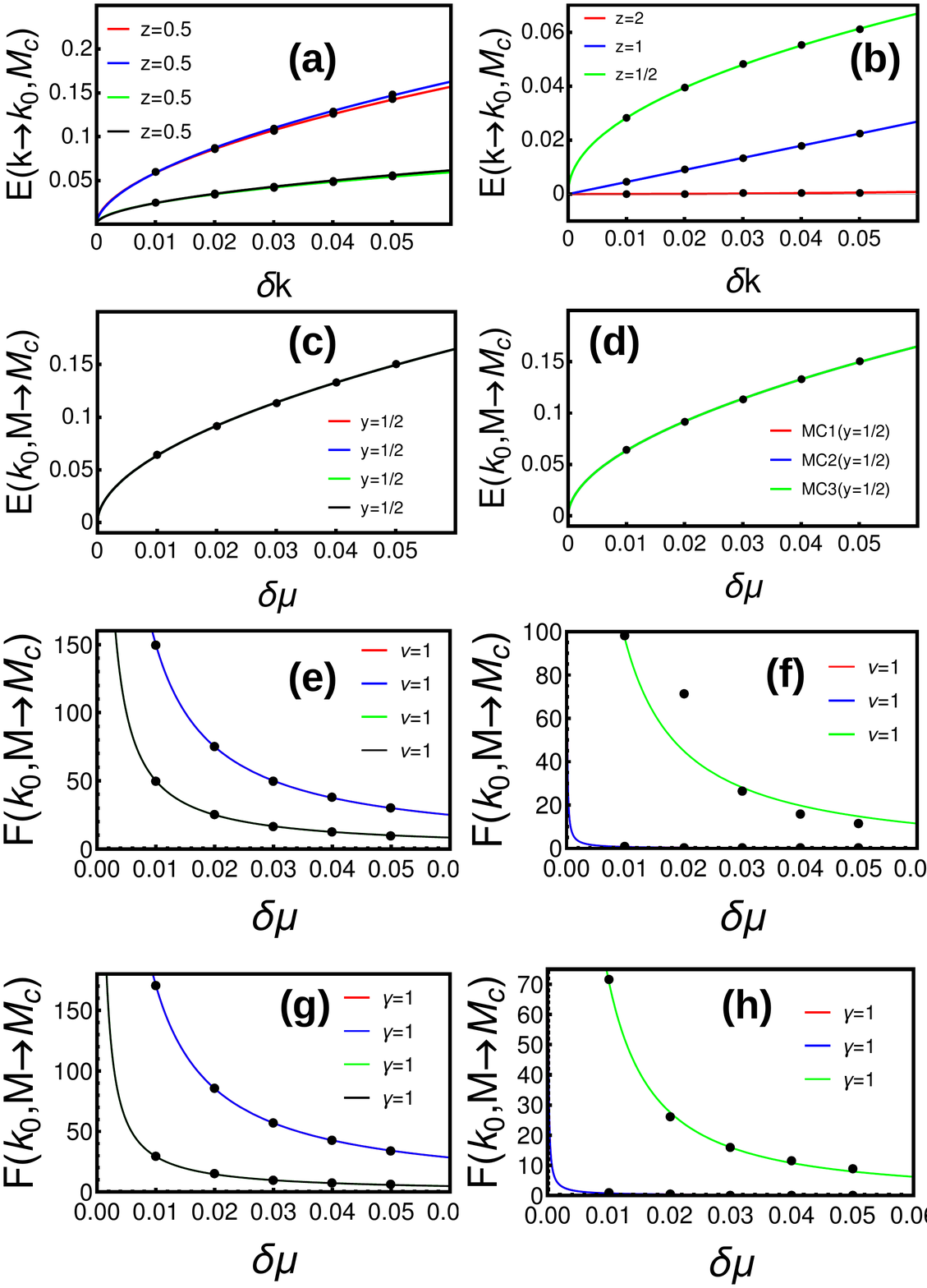}
	\caption{(Color online) Critical exponents through non-linear curve fitting method with the parameter space $t^{\prime}=1$ and $\delta=0.2$ respectively. The red, blue, green and black colors in left panel correspond to first ($\alpha=1.5,t_c=-1.15$), second ($1.5,-1.55$), third ($1.5,1.84$) and fourth ($1.5,0.44$) TQCLs respectively. The red, blue and green colors in right panel correspond to first (1,0.7), second (1,0.3) and third (0.53,0.5) multi-critical points respectively.(a-b) Dynamical critical exponent ($z$) (c-d) Crossover critical exponent ($y$) (e-f) Localization critical exponent ($\nu$). (g-h) Susceptibility critical exponent ($\gamma$)}
	\label{r22}
\end{figure}

In the vicinity of phase transition point, the physical quantities shows the diverging nature, which can be quantified by the critical exponents\cite{continentino2017quantum,continentino2020finite}. The set of critical exponents constitutes the universality class, which effectively categorizes the phase transitions based on their behavior. Here we analyze a few critical exponents (relevant to zero temperature TQPTs) and analyze their behavior.\\\\
\textit{The dynamical critical exponent $(z)$}  defines the nature of dispersion and can be calculated by expanding Eq.~\ref{endisp} around gap closing point $k_0$\cite{rufo2019multicritical}. i.e.,
\begin{equation}
E(k,\mathbf{M})=\sqrt{|\delta g|^2+A_1k+A_2k^2+A_4k^4+...},\label{sca}
\end{equation}
where the dominating terms among $A_1,A_2,A_4$ are real numbers which decides the nature of dispersion and corresponding $z$.\\
For the simplest longer-range model with two neighbors, the winding vectors are given by
\begin{eqnarray}
	\chi_x&=&t+\cos(k)+\frac{\cos(2k)}{2^{\alpha}},\nonumber\\
	\chi_y&=&\sin(k)+\frac{\sin(2k)}{2^{\alpha}}-i\delta,
\end{eqnarray}
with $t^{\prime}=1$. The criticality occurs at three values of $k$, i.e., $k=0,\pi$ and $\cos^{-1}(-2^{\alpha-1})$ which results in six exceptional points respectively.
Thus the model possesses six TQCLs 
\begin{eqnarray}
k&=&0\implies t=-\frac{1}{2^{\alpha }}+\delta -1, \hspace{0.2cm}
t= -\frac{1}{2^{\alpha }}-\delta -1\nonumber,\\
k&=&\pi\implies t=-\frac{1}{2^{\alpha }}+\delta +1,\hspace{0.2cm}
t=-\frac{1}{2^{\alpha }}-\delta +1\nonumber,\\
k&=&\cos^{-1}(-2^{\alpha-1})\implies t=\frac{1}{2^{\alpha }}+\delta,\hspace{0.2cm}
t= \frac{1}{2^{\alpha }}-\delta.\nonumber
\end{eqnarray}
 with red, blue, green, black, magenta and gray respectively (Fig~\ref{r2} a).\\
 Here we observe two different spectrum of real and absolute energy dispersion with different critical exponents. Due to the non Hermiticity, the energy dispersion need not to be symmetric around the gap closing values, which may result as different exponents around a same point\cite{arouca2020unconventional}.
 In our study, both real and absolute spectrum yield $z=1/2$ for all TQCLs except at multi-criticality (Fig~\ref{r22} a). At multi-criticality the real spectrum yield $z=2,1$ and 0.5 while (Fig~\ref{r22} b)absolute spectrum yield $z=1,1$ and 0.5 for MC1, MC2 and MC3 respectively (absolute spectrum is not showed here).\\
 The longer-range non-Hermitian SSH model contains both fractional and integer WNs, where the WN is given by Eq.~\ref{WN}, which physically means the encircling of the winding vectors around the origin of the pseudo-spin parameter space. Due to non-Hermiticity, there occurs two exceptional points instead of single Dirac point which act as center of pseudo-spin space. If the winding vectors encircle both the exceptional points equal times, then the WN is integer. If the winding vectors encircle either one of the exceptional point, or unequal number of time, then there occur fractional WN. \\
 In Fig.~\ref{r2} b, when winding vectors encircle both the EPs once (twice), we get $W=1(W=2)$ topological phase which is equivalent to Hermitian version of topological phases. If the winding vectors encircle either one of the exceptional point, it gives $W=1/2$ and encircling first (second) EP twice (once) while other only once (twice) give rise to $W=3/2$, which do not have any Hermitian counterpart. Under some special cases, winding vectors encircle one of the exceptional point even number of times and do not encircles the other, which also results in the integer WN. With the increasing decay parameter, the longer-range reduces to its short-range version at $\alpha=1$ through the intersection of TQCLs i.e., multi-critical points.\\\\
 \textit{Localization critical exponent $(\nu)$:} In a topological system, the zero energy modes are localized at the edges and protected by the bulk Bloch states.
 % and with the thermodynamic system size, the bulk shows extensive behavior and edge non-extensive behavior respectively.
%The Hamiltonian (Eq.\ref{Ham}) can be written for a simple extended chain (r=2) as,
%\begin{eqnarray}
%	H&=&\sum_{j=1}^{L-l}
%(t-\delta)c^{\dagger}_{j,a}c_{j,b}+(t+\delta)c^{\dagger}_{j,b}c_{j,a}\nonumber\\
%&+&\sum_{j=1}^{L-l}
% t^{\prime}(c^{\dagger}_{j+1,a}c_{j,b}+c^{\dagger}_{j,b}c_{j+1,a})\nonumber\\
% &+&\sum_{j=1}^{L-l}
% \frac{t^{\prime}}{2^{\alpha}}(c^{\dagger}_{j+2,a}c_{j,b}+c^{\dagger}_{j,b}c_{j+2,a}).
%\end{eqnarray}
 As a simple example, we consider Eq.\ref{Ham} with two neighbor couplings ($r=2$). For a semi-infinite system with $j=1,2,3...$, with wave function $\psi_n$, the energy relation yields\cite{yin2018geometrical}
 \begin{eqnarray}
 &&(t+\delta)\psi_{a,j}+t^{\prime}\psi_{a,j+1}+\frac{t^{\prime}}{2^{\alpha}}\psi_{a,j+2}=E_n\psi_{b,j}\nonumber\\
 &&(t-\delta)\psi_{b,j}+t^{\prime}\psi_{b,j-1}+\frac{t^{\prime}}{2^{\alpha}}\psi_{b,j-2}=E_n\psi_{a,j}.
 \end{eqnarray}
We construct the boundary condition with zero energy state ($E=0$) as
 \begin{eqnarray}
	&&(t+\delta)\psi_{a,j}+t^{\prime}\psi_{a,j+1}+\frac{t^{\prime}}{2^{\alpha}}\psi_{a,j+2}=0\nonumber\\
	&&(t-\delta)\psi_{b,j}+t^{\prime}\psi_{b,j-1}+\frac{t^{\prime}}{2^{\alpha}}\psi_{b,j-2}=0.\label{zero}
\end{eqnarray}
Here the boundary condition $\psi_{b,j}=0$ is true for every $n$. The ratio of nth localized zero eigenstate to that of first is given by\cite{continentino2017quantum,continentino2020finite}
\begin{equation}
	\delta\psi_n=\frac{\psi_n(E=0)}{\psi_1(E=0)}=\left| \frac{\delta g}{A_{1,2,4}}\right|^{n-1}.
\end{equation}
These states are ensured by the condition $e^{ik_0}=-\left( \frac{\delta g}{A_{1,2,4}}\right)$. Thus the $k_0$ is a complex number with $k_0=i\left( \frac{\delta g}{A_{1,2,4}}\right)$, where $A_{1,2,4}$ are the real numbers and dominating term among them decides the value of $k_0$. With the system size $n$, the localization of the eigenstates is given by,
\begin{equation}
	\delta\psi_n=e^{\frac{-(n-1)}{\xi}},
\end{equation}
where $\xi$ is the localization length with $\xi=\frac{A_n}{|\delta g|^{\nu}}\implies\xi=\xi_0|\delta g|^{-\nu}$, with $\xi_0$ as the natural length of the system. Here $A_n$ is the dominating term in Eq.~\ref{sca} and $\nu$ is the localization critical exponent.
  In non-Hermitian systems, the edge localization is highly influenced by the skin-effect which decides the bulk-edge correspondence. In our model, we get $\nu=1$ for both TQCLs and multi-critical points (Fig.\ref{r22} e,f).\\
To understand the number of localized edge modes at each end, we write Eq.\ref{zero} as
   \begin{eqnarray}
  &&(t+\delta)+t^{\prime}\lambda+\frac{t^{\prime}}{2^{\alpha}}\lambda^2=0,\nonumber\\
  &&(t-\delta)+t^{\prime}\lambda^{\prime}+\frac{t^{\prime}}{2^{\alpha}}\lambda^{\prime 2}=0.\label{z1}
  \end{eqnarray}
  with $\lambda=\frac{\psi_{a,j+1}}{\psi_{a,j}}$ and $\lambda^{\prime}=\frac{\psi_{b,j+1}}{\psi_{b,j}}$, corresponding to the left and right localized modes respectively. The quadratic form gives two roots
  \begin{equation}
  \lambda_{\pm}=\frac{-t^{\prime}\pm\sqrt{(t^{\prime})^2-4(\frac{t^{\prime}}{2^{\alpha}})(t\pm\delta)}}{2^{1-\alpha}(t^{\prime})}.
  \end{equation}
  The topological limit forms a unit circle, where the number of solutions less than unity represent the number of localized modes. The roots of $\lambda$ and $\lambda^{\prime}$ represent the left and right localized modes respectively. The combination of these two solution constitutes the phase diagram as mentioned in Ref.~\cite{yin2018geometrical}.\\\\
\textit{Crossover critical exponent $(y)$:} The scaling behavior near the TQCL is given by\cite{arouca2020unconventional},
\begin{eqnarray}
E(k\rightarrow k_0,\mathbf{M}=\mathbf{M}_c)&\propto& k^z,\nonumber\\
E(k=k_0,\mathbf{M}\rightarrow\mathbf{M}_c)&\propto& \left|\partial g \right|^{y},
\end{eqnarray}
where $y$ is called crossover or gap critical exponent with $y=z\nu$. This exponent gives the relation between $z$ and $\nu$ and can be calculated in the limit $E(k_0,\mathbf{M}\rightarrow\mathbf{M}_c)$. The energy gap at criticality behaves like $\Delta\propto|\delta g|^{z\nu}$, where $\delta g$ is the distance from the criticality. In our case, both real and absolute spectrum give $y=0.5$ for TQCLs as well as for multi-critical points (Fig.~\ref{r22} c,d).
\textit{Canonical critical exponent $(\alpha^*)$:} The concept of thermodynamic grand potential ($\Omega$) helps to understand the order of phase transition and corresponding critical exponent\cite{arouca2020unconventional,kempkes2016universalities}. The divergence or non-analyticity in the derivative of the grand potential signals the order of the corresponding transition. 
%The system with periodic boundary condition contains the information of bulk and open boundary contains both the bulk and boundary respectively. 
For a system with periodic boundary condition at zero temperature, the grand potential density is given by,
\begin{equation}
	\omega=\frac{\Omega}{N}=-\int_{-\pi}^{\pi}E(k,\mathbf{M})dk,\label{density}
\end{equation}
where $N$ is the system size. The above integral is 
%divergent for all $\alpha<1$ and finite for 
convergent for all positive $\alpha$, which makes the grand potential as independent of system size in this limit. 
The order of the transition can be calculated through the derivative of the grand potential. The transition can be nth order if the nth derivative of grand potential shows a cusp or spike against the parameter and the (n+1)th derivative shows a discontinuity\cite{malard2020multicriticality,chen2008intrinsic}.
 \begin{figure}[H]
	\centering
	\includegraphics[width=\columnwidth,height=8cm]{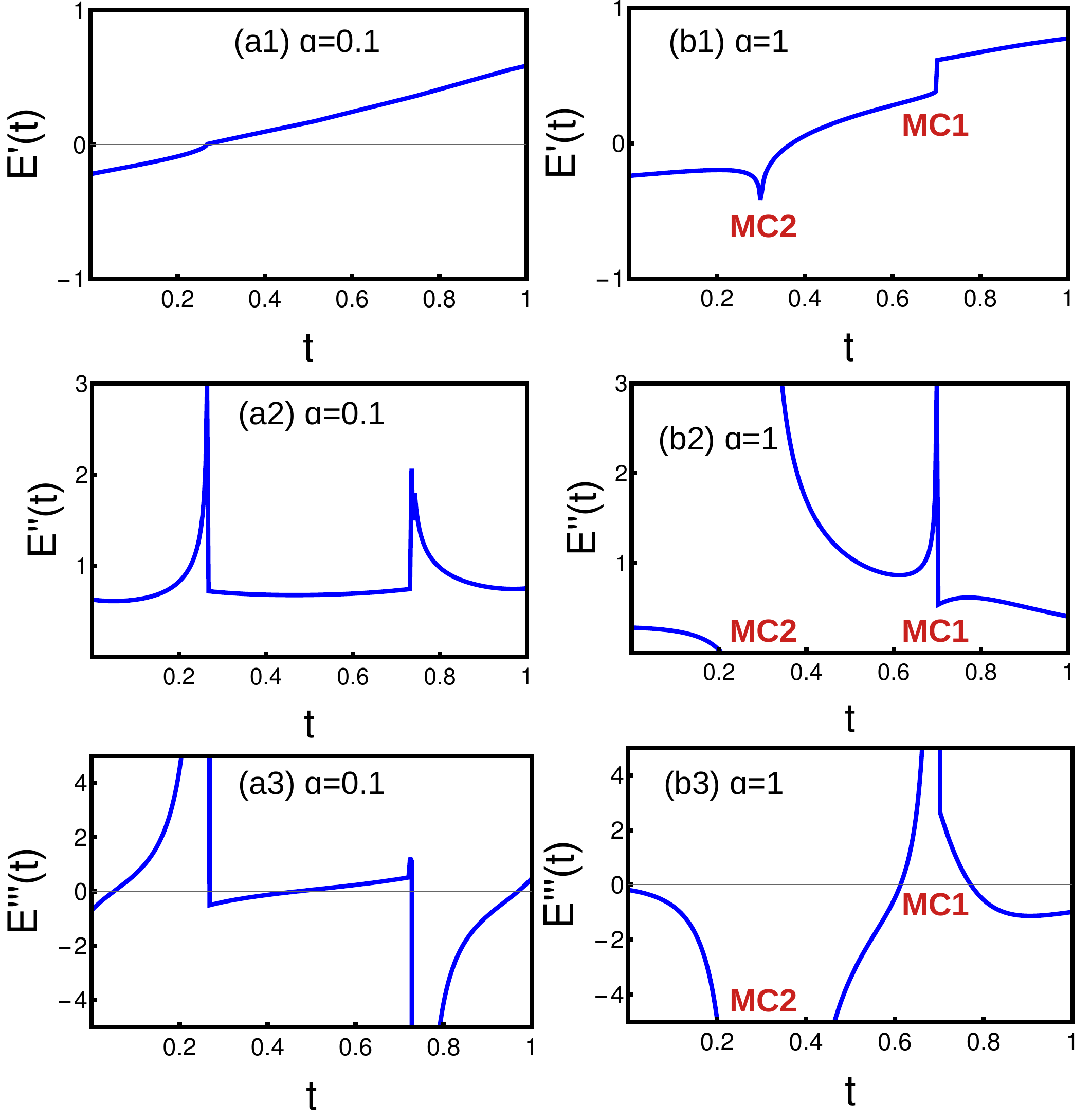}
	\caption{(Color online) Derivatives of ground state energy for $k=\pi$ critical line at different values of $\alpha$. (a1-a3) Normal critical lines showing cusp for second order derivatives. (b1-b3) Behavior of multi-critical points, where MC1 and MC2 show cusp for second and first order derivatives respectively.}
	\label{fig4}
\end{figure}
In Fig.~\ref{fig4}, we study $k=\pi$ critical line (t=1-$\frac{1}{2^{\alpha}}\pm\delta$) under two different configurations. For $\alpha=0.1$ the second derivative with respect to $t$ shows a cusp at criticality followed by a discontinuous third order derivative. At multi-criticality ($\alpha=1$), the first order derivative shows cusp at MC2 and a continuous curve at MC2. The second order derivative shows discontinuity at MC2 and a cusp at MC1 respectively. This observation signals that, the multi-criticalities MC1 and MC2 exhibit different nature of transitions. To understand this in detail we study the hyperscaling relation.\\
 Here, the grand potential ($\omega$) contains two parts, i.e., regular and singular parts. 
The singular part of the grand potential density scales as\cite{arouca2020unconventional}
\begin{equation}
	\omega_{singular}\propto|g|^{2-\alpha^*},
\end{equation}
where $\alpha^*$ is the canonical critical exponent and $2-\alpha^*$ signals the order of transition.
These critical exponents are connected by the relations called scaling laws which governs the phase transitions. One such relation is the Josephson's hyperscaling relation given by\cite{continentino2017quantum}
\begin{equation}
	2-\alpha^*=\nu(z+d),\label{hyp}
\end{equation}
where $z+d$ is the effective dimension.
 This law connects the dimensionality, dynamical and correlation critical exponents with order of transition through hyper-scaling relation. Thus, through hyper-scaling relation (Eq.\ref{hyp}), 
\begin{eqnarray}
	2-\alpha^*&=&1(1/2+1)=3/2\hspace{0.5cm}\text{for TQCLs},\nonumber\\
	2-\alpha^*&=&1(2+1)=3\hspace{0.5cm}\text{for MC1},\nonumber\\
	2-\alpha^*&=&1(1+1)=2\hspace{0.5cm}\text{for MC2},\nonumber\\
	2-\alpha^*&=&1(1/2+1)=3/2\hspace{0.5cm}\text{for MC3}.
\end{eqnarray}
% To get the scaling of the singular part of the energy density, we have to scale the relation with that of the correlation length. i.e.,$\omega_{singular}\propto g^{2-\alpha^*}g^{-\nu}$.
 %\textcolor{red}{Something more is required with a figure on energy density and scaling}
which clearly signals the difference between the order of transitions among normal criticality and multi-criticality. The fraction order of transitions are the interesting observations in non-Hermitian systems, which are different than their corresponding Hermitian counterparts. In Ref.~\cite{arouca2020unconventional}, the authors have tried to analyze this issue by scaling the singular part with the factor $\xi\propto|\delta g|^{-\nu}$. As the grand potential has a contribution from system size $L$, every distance is divided by the correlation length. This has showed that,  for the lower range of $\delta g$ the singular part of grand potential scales as $\delta g^{1/2}$, whereas for the higher range it scales as $\delta g^1$. This is the reason behind the difficulty to decide whether the transitions are first order or second order. However, in our model also we find similar observation at normal criticality, but multi-criticality remains interesting and merits more attention in this regard.
\subsection{Criticality and renormalization group approach}
\begin{figure}[H]
	\centering
	\includegraphics[width=\columnwidth,height=10cm]{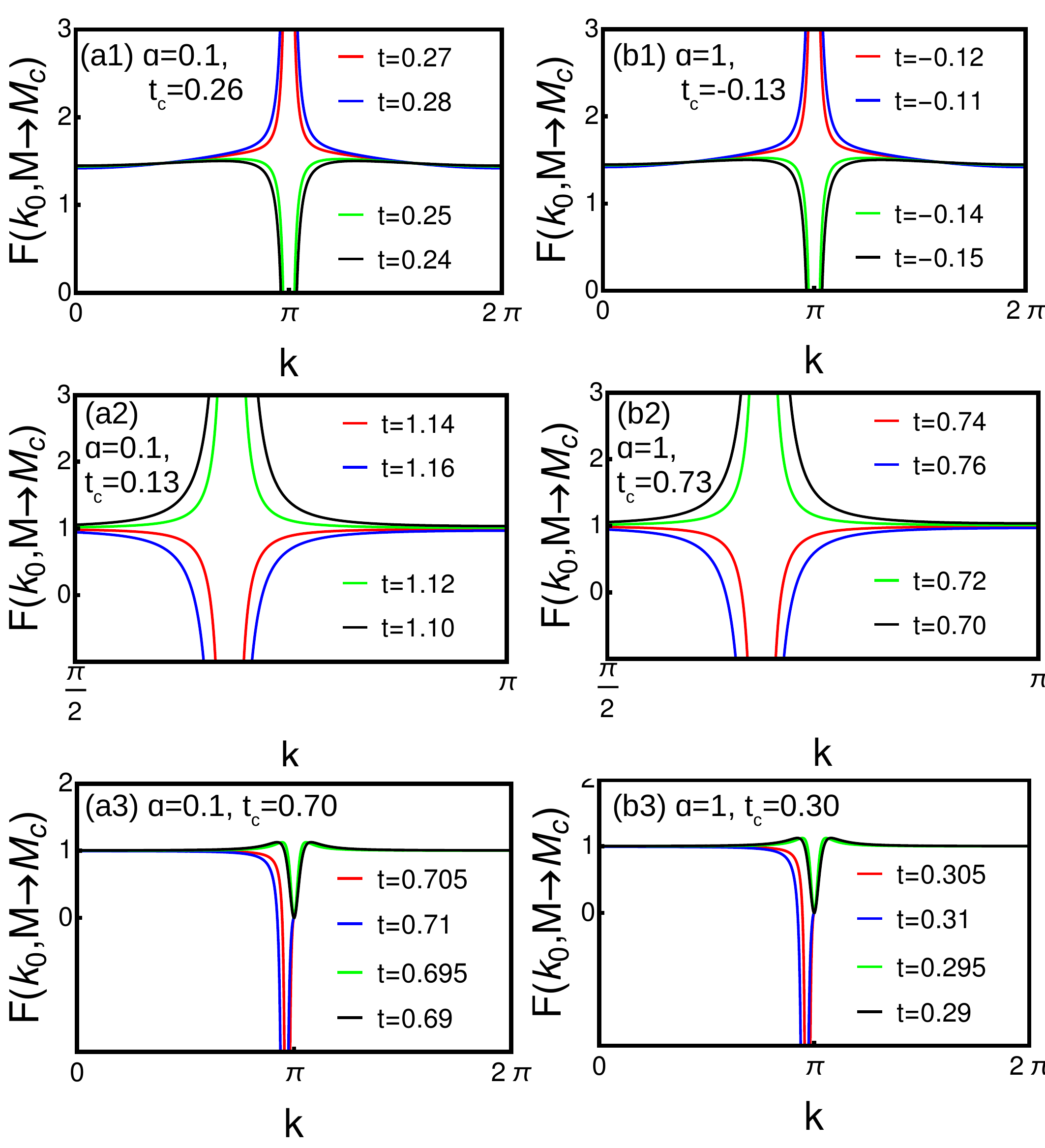}
	\caption{(Color online) Behavior of curvature function around HSP, non-HSP and fixed points respectively. The left panels represent criticality $t=1-\frac{1}{2^{\alpha}}-\delta$ ($F1(k,\mathbf{M})$ parameter space) and the right represent $t=1-\frac{1}{2^{\alpha}}+\delta$ ($F2(k,\mathbf{M})$ parameter space) respectively.}
	\label{cfr}
\end{figure}

Due to the non-Hermitian effects, the TQPT occurs through the exceptional points at $(0, \chi_y^{im})$ and $(0, -\chi_y^{im})$ instead of the Dirac points. The WN is given by Eq.~\ref{WN}, with
\begin{eqnarray}
	F_1(k,\mathbf{M})&=&\partial_k\phi_1=\frac{(\chi_x^{re}-\chi_y^{im})\partial_k\chi_y^{re}-\chi_y^{re}\partial_k(\chi_x^{re}-\chi_y^{im})}{(\chi_x^{re}-\chi_y^{im})^2+(\chi_y^{re})^2},\nonumber\\
	F_2(k,\mathbf{M})&=&\partial_k\phi_2=\frac{(\chi_x^{re}+\chi_y^{im})\partial_k\chi_y^{re}-\chi_y^{re}\partial_k(\chi_x^{re}+\chi_y^{im})}{(\chi_x^{re}+\chi_y^{im})^2+(\chi_y^{re})^2}.\nonumber\\\label{cf}
\end{eqnarray}
Here we deal with an interesting quantity called curvature function (CF). CF is the quantity whose integral over the Brillouin zone gives the topological invariant, like Berry connection, Berry curvature and Pfaffian wave function\cite{kumar2020multi,chen2019topological}. In 1D models, we consider Berry connection as our CF which is an gauge dependent quantity. The CF exhibits non-analyticity at the criticality and yields ill-defined topological invariant at these points. In case of non-Hermitian systems, exceptional points are the points of non-analyticity. Based on the behavior in the vicinity of exceptional points, CF can be classified as following.
\begin{enumerate}
\item High symmetry points (HSP): When the CF exhibits a symmetric nature around the gap closing point $k_0$, i.e.,$F(k_0+\delta k,\mathbf{M})=F(k_0-\delta k,\mathbf{M})$ is called a HSP. Here the Lorentzian peak is is symmetric under condition $k_0=-k_0$. The CF diverges as one approach $k_0$ from one direction $\mathbf{M}\rightarrow\mathbf{M}_c^+$ and flips the direction as one moves away $\mathbf{M}\rightarrow\mathbf{M}_c^-$ from $k_0$ with same diverging curve. This can be represented as fixed peak with varying height (Fig.~\ref{cfr} a1,b1).
\item Non-High Symmetry points (non-HSP): In this case, as one approaches criticality, there occurs diverging peak with different $k_0$ values. Here one can observe the flipping of peaks but the Lorentzian peak keep moving along the BZ (Fig.~\ref{cfr} a2,b2).
\item Fixed point (FP): Generally this kind of behavior is observed at the multi-criticality where HSP and non-HSP meet together\cite{kartik2021topological,kumar2020multi,kumar2021topological}. Here the CF shows a diverging peak as $\mathbf{M}\rightarrow\mathbf{M}_c^+$ from one side, and at $k_0$ we observe a peak with diminished height. As we go away $\mathbf{M}\rightarrow\mathbf{M}_c^-$ we observe a peak with constant height with increasing width. The multi-critical points MC1 and MC2 show the FP nature (Fig.~\ref{cfr} a3,b3).
\end{enumerate}
\textit{Susceptibility critical exponent $(\gamma)$:} Here the CF exhibits non-analytic property at the exceptional points, which gives the idea of critical scaling near these points\cite{chen2019topological}. In present case, we consider Berry connection as our CF, which is a gauge dependent quantity. Without loss of generality, we write CF in Ornstein-Zernike form as,
\begin{equation}
	F(k_0+\delta k,\mathbf{M})=\frac{F(k_0,\mathbf{M})}{1\pm\xi^2\delta k^2+\xi^4\delta k^4},
\end{equation}
where $k_0$ is a gap closing point, $\delta k$ is a small deviation from gap closing and $\xi$ is the characteristic scale respectively. At $k=k_0$ and $\mathbf{M}\rightarrow\mathbf{M}$, the CF diverges signaling a TQPT where the characteristic length vanishes.
\begin{equation}
	lim_{\mathbf{M}\rightarrow\mathbf{M}_c^+}F(k_0,\mathbf{M})=-lim_{\mathbf{M}\rightarrow\mathbf{M}_c^-}F(k_0,\mathbf{M})=\pm\infty.
\end{equation}
and CF exhibiting an interesting nature
\begin{equation}
F(k_0,\mathbf{M}^{\prime})=F(k_0+\delta k,\mathbf{M}),\label{CRG}
\end{equation}
which gives the idea of renormalization group study near criticality.
\begin{figure}[H]
	\centering
	\includegraphics[width=\columnwidth,height=12cm]{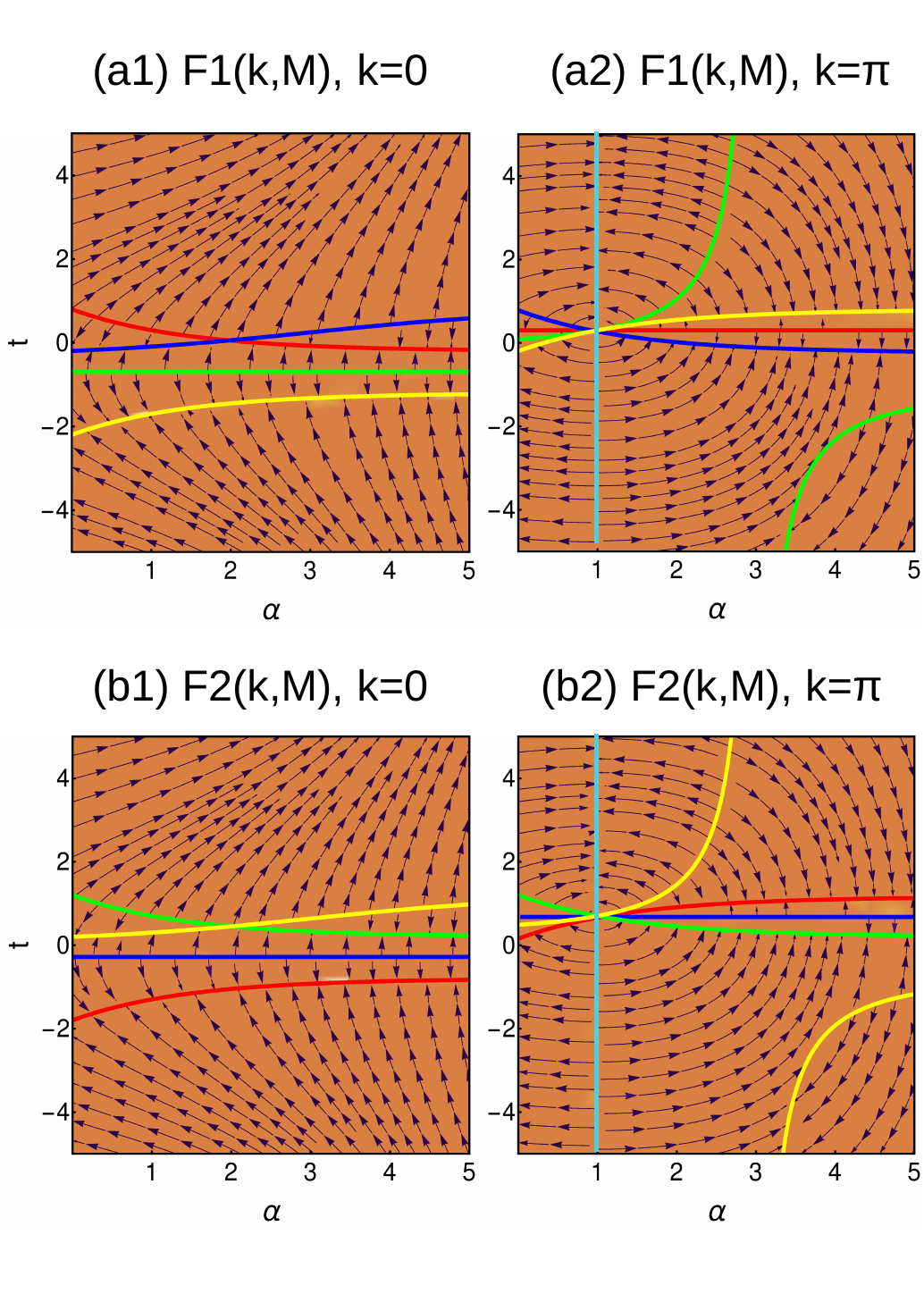}
	\caption{(Color online) CRG flow diagram for non-Hermitian longer-range SSH chain with two interacting neighbors. (a) Flow for $F1(k,\mathbf{M})$ at $k=0$. (a) Flow for $F1(k,\mathbf{M})$ at $k=\pi$. (a) Flow for $F2(k,\mathbf{M})$ at $k=0$. (a) Flow for $F2(k,\mathbf{M})$ at $k=\pi$. Here the colored lines represent critical and fixed configurations. For detailed derivation, refer Appendix\ref{appb}.}
	\label{CRGr2}
\end{figure}
 The divergence of CF and characteristic length gives the critical exponent $F(k_0,\mathbf{M})\propto|\mathbf{M}-\mathbf{M}_c|^{-\gamma}$ and $\xi(k_0,\mathbf{M})\propto|\mathbf{M}-\mathbf{M}_c|^{-\nu}$, where $\gamma$ and $\nu$ are the susceptibility and characteristic critical exponents respectively, where the susceptibility exponent explains the quality of the transition to be topological. Here we get $\gamma=1$ for both TQCLs and multi-critical points.\\
Expanding Eq~\ref{CRG} up to a leading order, we get CRG equation as\cite{chen2019topological}, 
\begin{equation}
	\frac{d\mathbf{M}}{dl}=\frac{(\frac{1}{2})\frac{d^2F(k,\mathbf{M})}{d k^2}|_{k=k_0}}{\frac{dF(k_0,\mathbf{M})}{d\mathbf{M}}},\label{rg}
\end{equation}
where $\left| \frac{d\mathbf{M}}{dl}\right|\rightarrow\infty$ defines criticality and $\left| \frac{d\mathbf{M}}{dl}\right|\rightarrow0$ defines fixed point configuration respectively. In our model, at $k=\pi$ (both $F1(k,\mathbf{M})$ and $F2(k,\mathbf{M})$), there occurs a superposition of all critical and fixed lines at $\alpha=1$. Here we find a whirlpool kind of behavior where the flow lines encircle a  single point (Fig.~\ref{CRGr2}). At this point, $\frac{d^2F}{dk^2},\frac{dF}{d\alpha},\frac{dF}{d\mu}\rightarrow0$ and $\frac{d\alpha}{dl},\frac{d\mu}{dl}\rightarrow\frac{0}{0}$. Interestingly at this point, we observe a different class of UCCE through multi-critical points. On the other hand, we do not observe any such behavior for $k=0$ criticality. (For detailed derivation, refer Appendix\ref{appb})
	\begin{table}[H]
	\begin{center}

		\begin{tabular}{ |c|c|c|c|c|c|c| } 
			\hline
			Model &$\hspace{0.2cm}k\hspace{0.2cm}$&$\hspace{0.2cm}z\hspace{0.2cm}$&$\hspace{0.2cm}\nu\hspace{0.2cm}$& $\hspace{0.2cm}\gamma_{1,2}\hspace{0.2cm}$&$\hspace{0.2cm} y\hspace{0.2cm}$&
			$2-\alpha^*$\\ 
			\hline
			\hline
			Short-range model
			&0& 0.5 & 1&1& 0.5&1.5\\
			
			&$\pi$& 0.5 & 1&1& 0.5&1.5\\
			\hline
			Longer-range model&0& 0.5 & 1&1& 0.5&1.5\\
			($r=$ Even and $\alpha=1$)	
			
			&$\pi$& (2,1) & 1&1& 0.5&(3,2)\\
			\hline
			Longer-range model&0& 0.5 & 1&1& 0.5&1.5\\
			($r=$ Odd, $\forall\alpha$)	 &$\pi$& 0.5 & 1&1& 0.5&1.5\\
			
			($r=$ Even, $\alpha\neq1$)	&&&&&&\\
			\hline
			\hline
		\end{tabular}
		
	\end{center}
	\caption{A comparison of universality class of critical 
		exponents between the original Kitaev chain and the reduced 
		long-range Kitaev chain.}
	\label{uc}
\end{table}

\subsection{Significance of multi-criticality}
 Multi-criticalities are the intersection of (at least) two TQCLs and distinguish (at least) three different topological regimes. So far, multi-critical points acted as unique points, where we witnessed the breaking of Lorentz invariance, topological transition without bulk gap and as a point with zero entanglement entropy\cite{kumar2021topological}. There are category of multi-critical points based on the conformal field theory where we get different central charge. However, in case of non-Hermitian systems, multi-critical points may behave differently based on the nature of non-Hermitian factor. For example, recently there has been the observation on the vanishing of multi-criticality when the non-Hermiticity is introduced into an extended Kitaev chain through chemical potential\cite{rahs}. In our case, we have introduced non-Hermiticity through local imbalance in the hopping, which splits the Hermitian multi-critical points into two. We can observe that multi-critical points as the superposition of HSP and non-HSP with a fixed point behavior and belong to separate UCCE. This part is in similarity with Hermitian counterpart.
  \begin{figure}[H]
 	\centering
 	\includegraphics[width=\columnwidth,height=10cm]{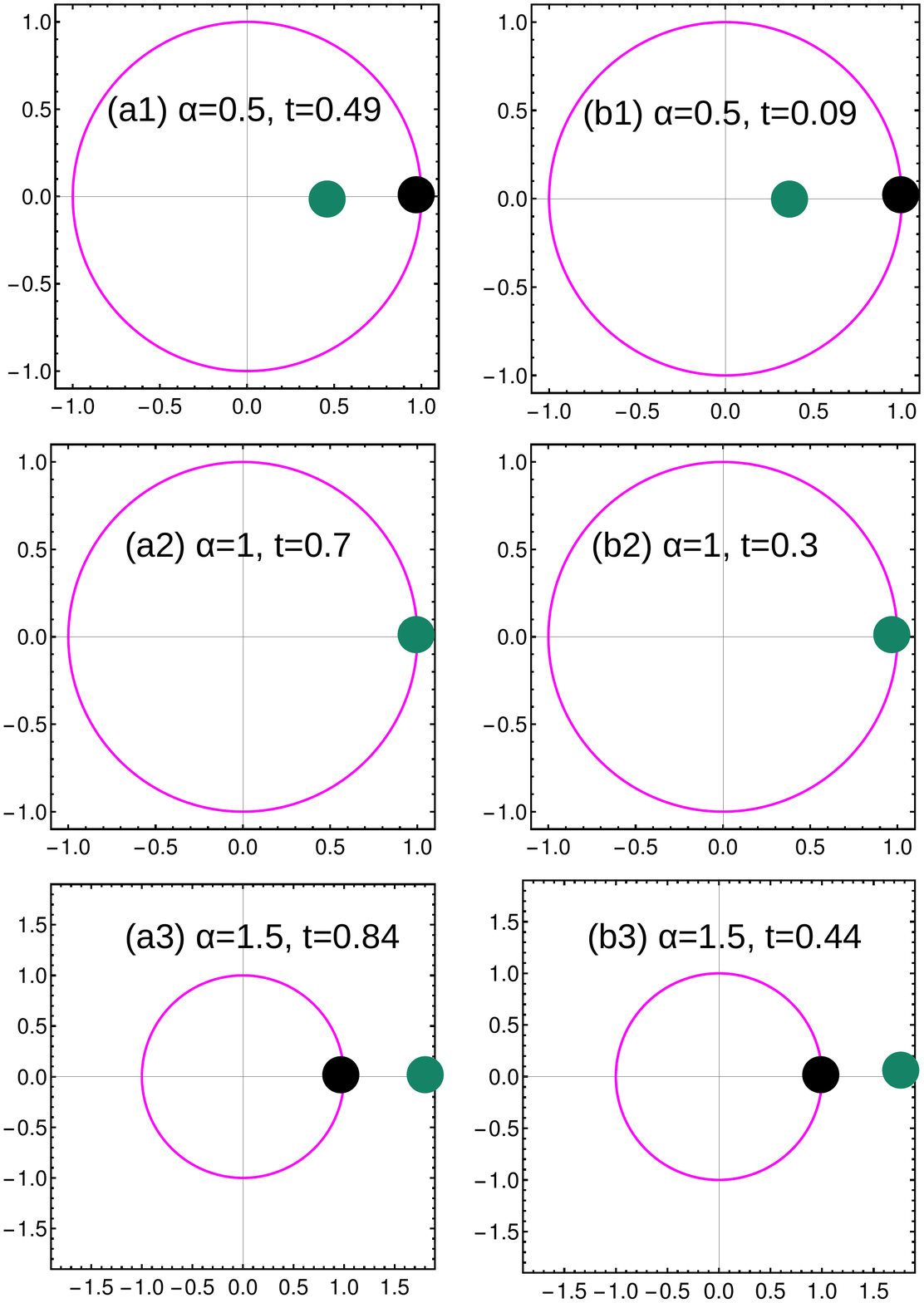}
 	\caption{(Color online) Localization of edge modes at criticality calculated through Eq.~\ref{z1}. The unit circle represent the topological limit. The solution less than, equal to and greater than unity represent the edge mode, criticality and no edge mode condition respectively. (a1-a3) Edge mode behavior at left end. (b1-b3) Edge mode behavior at right end.}
 	\label{com}
 \end{figure} 
The expansion of Eq.~\ref{endisp} around exceptional points $k_0$ gives the better understanding towards the underlying physics of multi-criticality. At normal criticalities, the dispersion behaves as $E(k\rightarrow k_0,\mathbf{M}_c)\propto k^{1/2}$. At multi-criticalities, the dispersion takes first ($E(k\rightarrow k_0,\mathbf{M}_c)\propto k^{1}$ for MC1) and second order ($E(k\rightarrow k_0,\mathbf{M}_c)\propto k^{2}$ for MC2) corrections as shown in Eq.\ref{sca} respectively. On the other hand, the multi-critical point MC3 is no different than the normal criticalities, as it shows similar behaviors of normal criticality. However, we observe similar cases in Hermitian counterpart, where the the instance is recognized as breaking of Lorentz invariance\cite{kumar2020multi}.\\
Exponent $z$ gives information about the massless condition of the system. In condensed matter, the energy dispersion at Fermi surface resembles the relativeistic energy form $E^2=c^2p^2+m^2c^4$. For gap closing points, the equation gives masless condition with $E^2=c^2p^2$. Under some cases, the equation takes correction form as, $E^2=c^2p^2+m^2c^4+Ap^4$ with $A$ as constant\cite{sadhukhan2020sonic}. Similar effects we observe at exceptional points. For normal criticalities, we get $E^2=\sqrt{AK}$ and for multi-criticalities, we obtain $E^2=\sqrt{Ak+Bk^2}$ and $E^2=\sqrt{Ak+Bk^4}$ respectively. This creates the difference in the nature of energy dispersion and thereby dynamical critical exponent.
 \textit{Transition among gapless phases:} Localization at criticality is an important phenomenon, which is important both from the perspective of theory and experimental aspects ~\cite{jones2019asymptotic,thorngren2020intrinsically,kestner2011prediction,cheng2011majorana,fidkowski2012majorana,sau2011number,kraus2013majorana,scaffidi2017gapless,jiang2018symmetry}, especially in longer-range models\cite{rahul2021majorana,kumar2020multi,verresen2019gapless,verresen2018topology,verresen2020topology,kumar2021topological}. In our model, the multi-critical points act as an unique point, which witnesses the topological transition among critical phases. Here MC1 and MC2 are present on the $k=\pi\rightarrow t=1\pm\delta+\frac{t^{\prime}}{2^{\alpha}}$ respectively. 
 Through Eq.\ref{z1}, we find the solutions of quadratic equation as shown in Fig.~\ref{com}. Throughout the critical line, one of the solution remains unity, which confirms the gapless condition. For $\alpha<1$, one of the solution is less than unity ($\lambda=0.4$), indicating a localization of edge  mode at criticality. At $\alpha=1$, the both the solutions are unity, indicating a transition among gapless phases through multi-criticality. For $\alpha>1$, one of the solution is greater than unity ($\lambda=1.8$), indicating the absence of localized modes. Hence, both MC1 and MC2 act as transition points among gapless phases. A similar phenomenon has been observed in the Hermitian counterpart of the similar model\cite{rahul2021majorana,kumar2020multi}. 
% \begin{equation}
%\frac{dE}{dk}=i\frac{\left(2^{1-\alpha }-1\right) \delta }{\sqrt{\left(2^{-\alpha }+t-1\right)^2-\delta ^2}}.
 %\end{equation}
\subsection{Staircase of topological transitions:}
Due to the feature of long-range coupling in Eq.\ref{Ham}, it is possible to achieve higher winding number with the increase of interacting neighbors, i.e., finite $r$. For the limit $r<L/2$, we can obtain higher WNs where the uppermost WN is $W=r$. This is an effective method to obtain more localized modes through static method while similar can be achieved through periodic driving which is dynamical in nature\cite{kartik2021topological,tong2013generating}. However the higher WNs are comparatively less stable and reduce to consecutive lower WNs with the increasing of $\alpha$. Here we find an interesting patter of transitions , i.e., a staircase of TQPTs. 
\begin{itemize}
\item Even-Even transition: When the number of neighbor is an even number (for $r>2$), the uppermost WN will be an even number and the transition occurs among only even WNs. 
\item Odd-Odd transition: When the number of neighbor is an odd number (for $r>2$), the uppermost WN will be an odd number and the transition occurs among only odd WNs. 
\item Even-Odd transition: When the number of neighbor is an even number (for $r>2$), the uppermost WN is an even number and with the increasing $\alpha$, the longer-range model reduces to short-range version. At the interface of short-range and longer-range, there occurs a transition between $W:2\rightarrow1$. For very small $\delta$, this can be clearly observed and with the increase in $\alpha$, the transition occurs like $W:2\rightarrow\frac{3}{2}\rightarrow1$. This transition occurs through the multi-critical points.
\end{itemize}
\begin{figure}[H]
	\centering
	\includegraphics[width=\columnwidth,height=14cm]{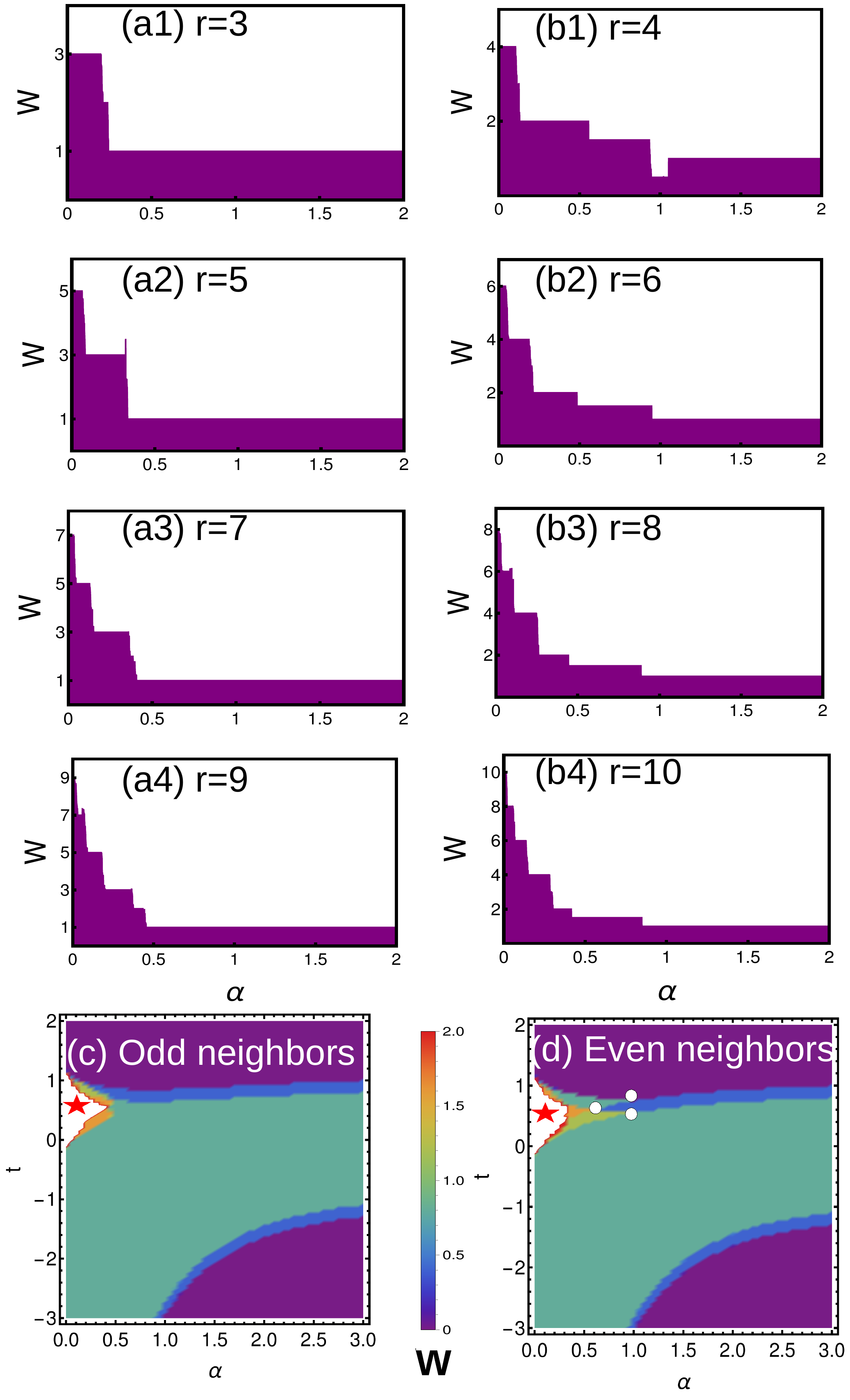}
	\caption{(Color online) Staircase of TQPTs with different number of interacting neighbors. For even (odd) '$r$', transitions occurs among only even (odd) winding numbers. Near short-range limit, we do not observe this behavior. The star mark indicates the numerical limitation of converging for higher WNs.}
	\label{staircase}
\end{figure}
Here we choose parameter space ($t=0.5, \delta=0.1$), where the increase in $\delta$ suppresses the uppermost WN and creates fractional WNs. For the even $r$, the fractional WNs are observable with small $\delta$ and large $\alpha$, whereas the similar effect is not visible for odd $r$. The interface of short-range and longer-range occurs through multi-critical points as shown in Fig.~\ref{staircase}(d). These multi-critical points can also be recognized by the CRG method also (Not shown here).
\section{Long-range}\label{sec3}
\begin{figure*}
	\centering
	\includegraphics[width=14cm,height=12cm]{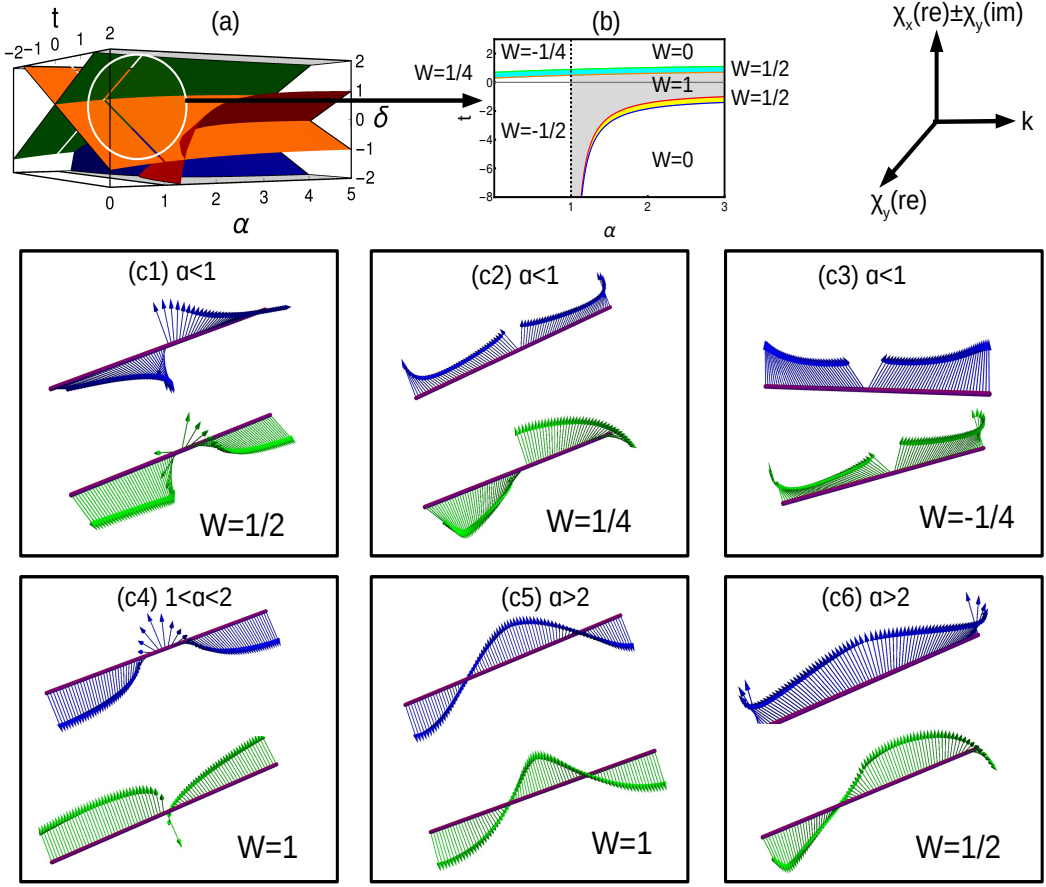}
	\caption{(Color online) (a) Phase diagram of long-range model in a 3D representaion. (b) 2D representation of phase diagram. (c1-c6) Behavior of winding vectors in long-range non-Hermitian SSH chain. The blue and green winding vectors represent the F1(k,$\mathbf{M}$) and F2(k,$\mathbf{M}$) parameter spaces respectively. The combined effect determines the topological index of the system.}
	\label{long}
\end{figure*}
With the introduction of infinite number of neighbors ($r\rightarrow\infty$), Eq.\ref{Ham} represents long-range non-Hermitian SSH chain, where the pseudo-spin parameters exhibit the polylogarithmic nature\cite{vodola2014kitaev}, i.e.,
 \begin{eqnarray}
 \chi_x(k)&=&t+t^{\prime}\left( \frac{Li_{\alpha}[e^{i k}]+Li_{\alpha}[e^{-i k}]}{2}\right),\nonumber\\
 \chi_y(k)&=&t^{\prime}\left( \frac{Li_{\alpha}[e^{i k}]-Li_{\alpha}[e^{-i k}]}{2i}\right)-i\delta.\label{pol}
\end{eqnarray}
Expansions of polylogarithmic function around a gap closing point $k_0$ is given by\cite{olver2010nist},
\begin{equation}
	Li_{\alpha}[e^{ik}]=\Gamma[1-\alpha](-ik)^{\alpha-1}+\sum_{n=0}^{\infty}\frac{\zeta[\alpha-n]}{n!}(ik)^n.\label{ref}
\end{equation}
For the detailed study, we can expand the polylogarithmic function around $k=0$ as
\begin{eqnarray}
	\chi_x&=&t+t^{\prime}(\Gamma[1-\alpha](k)^{\alpha-1}\sin\left( \frac{\pi\alpha}{2}\right) \nonumber\\
	&+&\sum_{n=0}^{\infty}\frac{\zeta[\alpha-n]}{n!}(k)^n\cos\left( \frac{\pi n}{2}\right) ),\nonumber\\
	\chi_y&=&t^{\prime}(\Gamma[1-\alpha](k)^{\alpha-1}\cos\left( \frac{\pi\alpha}{2}\right) \nonumber\\
	&+&\sum_{n=0}^{\infty}\frac{\zeta[\alpha-n]}{n!}(k)^n\sin\left( \frac{\pi n}{2}\right) )-i\delta.\label{poly}
\end{eqnarray}
We analyze the properties of momentum space to characterize the non-Hermitian long-range SSH chain.
\subsection{Energy dispersion, Fermi velocity, GSE density, critical exponents:}
\begin{figure}[H]
	\centering
	\includegraphics[width=\columnwidth,height=9cm]{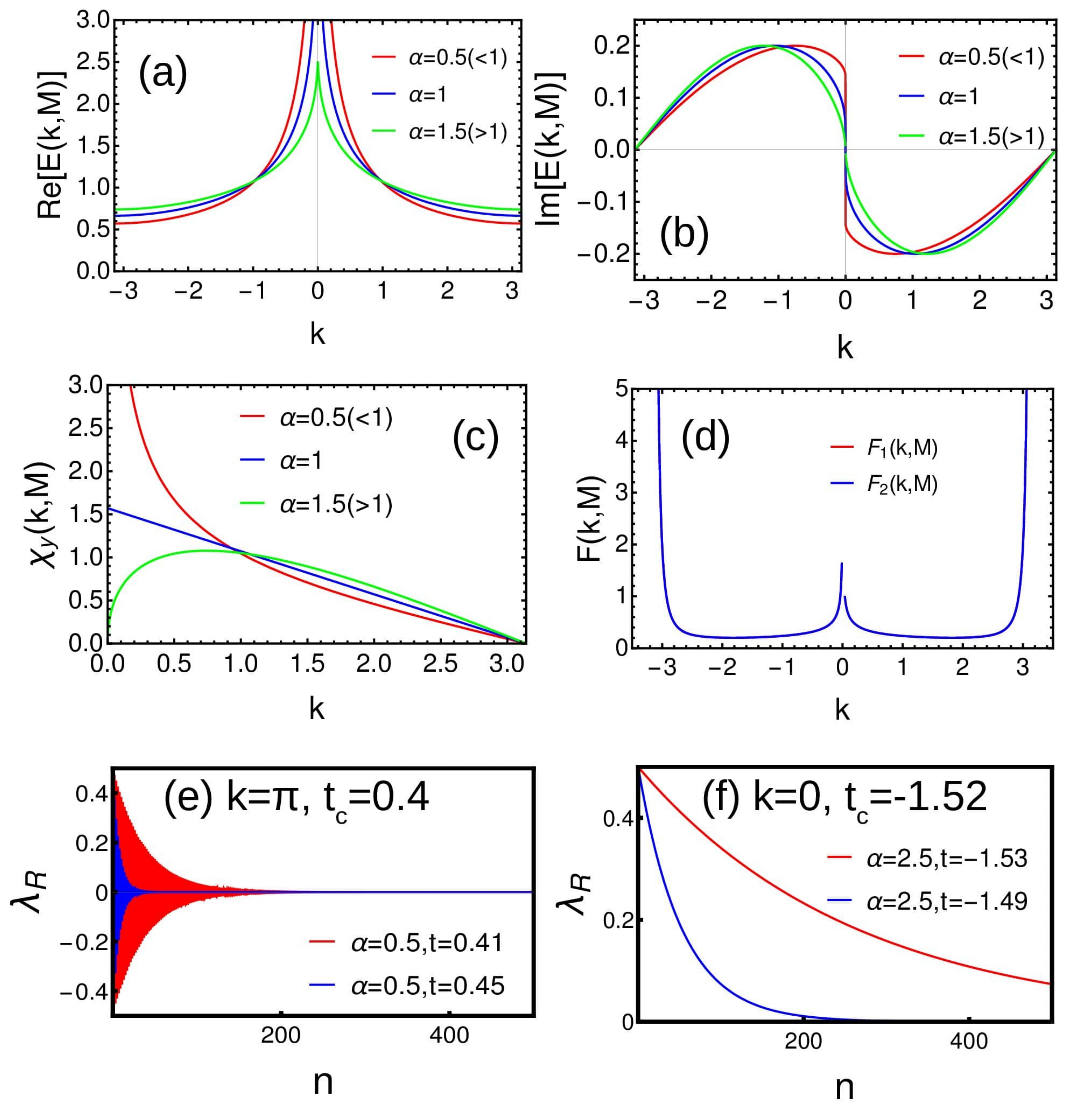}
	\caption{(Color online) Properties of non-Hermitian long-range SSH chain. (a) Real part of energy spectrum where the spectrum is divergent for $\alpha<1$ and representation of topological transition at $\alpha=1$ without gap closing. (b) A sharp variation of imaginary part of spectrum at $\alpha=1$ interface. (c) Divergence of $\chi_y$ term for $\alpha<1$ region due to the polylogarithmic behavior. (d) Behavior of curvature function for $\alpha<1$ region. The divergences at $k=\pi$ and 0 is due to the criticality and removable singularity respectively. (e) Decay of Wannier state correlation function around $k=\pi$. The nature is similar $\forall\alpha$. (f) Decay of correlation function around $k=0$, only for $\alpha>2$.}
	\label{prop}
\end{figure}

The quasi-particle energy dispersion is given by Eq.~\ref{endisp}, where the band gap vanishes for $k=0$ and $\pi$ respectively. Due to the polylogarithmic nature, the band gap closes for all values of $\alpha$ at $k=\pi$ and only $\alpha>1$ at $k=0$. The parameter space possesses two exceptional points and four TQCLs corresponding to two values of $k_0$. i.e.,
\begin{enumerate}
	\item $t=\delta -\text{Li}_{\alpha }(1)$ for $k=0$ for $\alpha>1$,
	\item $t=-\delta -\text{Li}_{\alpha }(1)$ for $k=0$ for $\alpha>1$,
\item	$t=\delta -\text{Li}_{\alpha }(-1)$ for $k=\pi$ and $\forall\alpha$,
\item	$t=-\delta -\text{Li}_{\alpha }(-1)$ for $k=\pi$ and $\forall\alpha$.
\end{enumerate}
Here the pseudo-spin parameters go as $\sin(k)$ instead of $\sin(lk)$ which result in the single time encircling of the origin.
% When the winding vectors encircle the centers of both the  parameter spaces $P1$ and $P2$, 
Thus the phase diagram is given by Fig.~\ref{long}(a,b). \\\\
\textit{Behavior of winding vectors:}
Due to the long-range effect, the winding vectors exhibit a discontinuity at $k=0$ for $\alpha<1$ which give rise to fractional winding number. Thus the region $\pm\delta -\text{Li}_{\alpha }(1)<\mu<\pm\delta -\text{Li}_{\alpha }(-1)$ for $\alpha>1$ gives topological (Hermitian and non-Hermitian) phase where the winding vectors encircle the origin of the parameter space without any discontinuity. Due to the polylogarithmic behavior, we find a less population of winding vectors in the region $1<\alpha<2$, technically which do not affects the winding number. In the region  $\delta -\text{Li}_{\alpha }(1)<\mu<-\delta -\text{Li}_{\alpha }(-1)$ for $\alpha>1$, we find that both set of winding vectors encircle the origin of each parameter space which yield $W=1$. 
The region $\mu<\pm\delta -\text{Li}_{\alpha }(-1)$ for $\alpha<1$ gives topologically ill-defined (fractional) phase where winding vectors corresponding to both the parameter space shows a discontinuity at $k=0$. If both set of vectors exhibit half encirclement around their corresponding origin, then it yield $W=1/2$. If any one exhibits this behavior, it yields $W=\pm1/4$ where the sign indicates the direction of encircling. On the other hand, due to the imbalance in intracell hopping, there arise fractional WN for 
$-\delta -\text{Li}_{\alpha }(1)<\mu<\delta -\text{Li}_{\alpha }(1)$ and $-\delta -\text{Li}_{\alpha }(-1)<\mu<\delta -\text{Li}_{\alpha }(-1)$ for $\alpha>1$. These are the non-Hermitian topological phases and do not posses any Hermitian counterpart. Here any one set of the winding vector encircle the origin by yielding $W=1/2$.
Extended winding vectors are given by
\begin{eqnarray}
\chi_x(ext)&=&\pm\delta +\frac{\text{Li}_{\alpha }(\exp (-i k))+\text{Li}_{\alpha }(\exp (i k))}{2}+t,\nonumber\\
\chi_y(ext)&=&\frac{\text{Li}_{\alpha }(\exp (i k))-\text{Li}_{\alpha }(\exp (-i k))}{2 i}.
\end{eqnarray}
The behavior of extended winding vectors are presented in Fig.~\ref{long}(c1-c6).\\\\
\textit{Momentum space behavior:} Due to the polylogarithmic nature, in the limit $k\rightarrow0$, the $\chi_y$ term shows a divergence for $\alpha<1$ and converges towards zero for $\alpha>1$ and a discontinuity at $\alpha=1$ respectively(Fig.~\ref{prop}c). This reflects in the behavior of CF and energy dispersion. For $\alpha<1$ the CF shows a divergence (Fig.~\ref{prop}d) at $k=0$ which leads to the emergence of the fractional WN. The corresponding energy spectrum shows a gapped region with diverging bands(Fig.~\ref{prop}a).The transition from $\alpha<1$ to $\alpha>1$ occurs at $\alpha=1$, where we observe topological transition without gap closing $(W: -1/4\rightarrow0,1/4\rightarrow1/2, W:-1/2\rightarrow1)$. This is an interesting behavior which can be observed in long-range models.
\textit{Curvature function and Wannier state correlation function:} Substituting the values of 
Eq.~\ref{pol} into Eq.~\ref{cf} we  get the CF in Ornstein-Zernike form as
\begin{eqnarray}
F(k,\mathbf{M})\propto\frac{\frac{A k^{\alpha-2}}{B k^{\alpha-1}}-\frac{Ck^{2\alpha-3}}{Dk^{2\alpha-2}}}{1+(\frac{Ek^{\alpha-1}}{Fk^{\alpha-1}})^2}.
\end{eqnarray}
There are three possible cases,
\begin{itemize}
	\item When $\alpha<1$, the term $k^{\alpha-2}$ dominates and $F(k,\mathbf{M})\rightarrow\infty$ as $k\rightarrow0$, irrespective of $\mathbf{M}\rightarrow\mathbf{M}_c$.
	\item When $1<\alpha<2$, the term $k^{\alpha-2}$ dominates and $F(k,\mathbf{M})\rightarrow\infty$ as $k\rightarrow0$, irrespective of $\mathbf{M}\rightarrow\mathbf{M}_c$.
	\item When $\alpha>2$, again the term $k^{\alpha-2}$ dominates and $F(k,\mathbf{M})\rightarrow\infty$ as $k\rightarrow0$ with $\mathbf{M}\rightarrow\mathbf{M}_c$.
\end{itemize}
Thus we can define the CF in the Ornstein-Zernike form and corresponding critical exponents only for $\alpha>2$ around $k=0$ region\cite{sadhukhan2021there}. For $k=\pi$, the CF can expressed Ornstein-Zernike form for all values of $\alpha$.\\
Wannier state correlation function is the quantity, which discusses the phase transition among different phases\cite{kumar2020multi,chen2019topological,kumar2021topological}. Conceptually, it is the overlap between the Wannier state centering at the origin and that of distance $R$. i.e.,
\begin{eqnarray}
\langle r|R\rangle=W_n(r-R)
\end{eqnarray}
with \begin{eqnarray}
|R_n\rangle=\frac{1}{N}\sum_{k}e^{ik.(r-R)}|u_k\rangle\nonumber
\end{eqnarray}
(the form of $|u_k\rangle$ is expressed in Appendix \ref{appa}.) \\
Mathematically, Wannier states can be expressed as Fourier transform of the CF, i.e.,
\begin{equation}
\lambda_R=\int\frac{dk}{2\pi}e^{ik.R}F(k,\mathbf{M}).
\end{equation}
In our case, we observe the decay of Wannier state correlation function around $k=\pi$ for all values of $\alpha$, with a critical exponent $\nu=1$ (Fig.~\ref{prop} e). For $k=0$, there occurs no correlation decay in the region $1<\alpha<2$, which results in the undefined critical exponent. For the region $\alpha>2$, we observe the decay of correlation function, which is very sharp near the criticality with an exponent $\nu=1$ (Fig.~\ref{prop} f).\\\\
\textit{Energy dispersion and Fermi velocity:} By substituting Eq.~\ref{pol} into Eq.~\ref{endisp}, we get energy dispersion. The first order derivative of the energy dispersion gives the Fermi velocity as
\begin{eqnarray}
	E(k,\mathbf{M})&=&\sqrt{Ak^{2\alpha-2}+Bk^{\alpha-1}+C},\nonumber\\
	\frac{dE_k}{dk}&=&\frac{Ak^{2\alpha-3}+Bk^{\alpha-1}+Ck}{\sqrt{A^2k^{2\alpha-2}+Bk^{\alpha}+C^2k^2}}.
\end{eqnarray}
where $A,B$ and $C$ are constants. For the values $\alpha<1$, the Fermi velocity is dependent on $\alpha$ as shown in Table~\ref{fermi}
\begin{table}[H]
	\begin{center}
		\begin{tabular}{|c|c|c|c|} 
			\hline
			Condition&Relation&$k=0$&$k=\pi$\\
			\hline
			\hline
			$\alpha<1$&$\frac{Ak^{2\alpha-3}}{Bk^{\alpha-1}}$&$v_F\rightarrow\infty$&$v_F(\mathbf{M}_c)\rightarrow0$\\
			$1<\alpha<1.5$&$Ak^{2\alpha-3}$&$v_F\rightarrow\infty$&$v_F(\mathbf{M}_c)\rightarrow0$\\
			$\alpha>1.5$&$Ck$&$v_F(\mathbf{M}_c)\rightarrow0$&$v_F(\mathbf{M}_c)\rightarrow0$\\
			\hline
		\end{tabular}
	\end{center}
	\caption{Behavior of Fermi velocity with respect to decay parameter $\alpha$. At $k=0$, the Fermi velocity vanishes as the system drives towards criticality only for $\alpha>1.5$}
	\label{fermi}
\end{table} 
Thus, as $k\rightarrow0$ we obtain a diverging energy spectrum and Fermi velocity for $\alpha<1$. For $1<\alpha<1.5$ we obtain gapless energy spectrum and a diverging Fermi-velocity as $k\rightarrow0$. For $\alpha>1.5$, we obtain vanishing Fermi velocity as the system drives towards criticality.
\begin{figure}[H]
	\centering
	\includegraphics[width=\columnwidth,height=12cm]{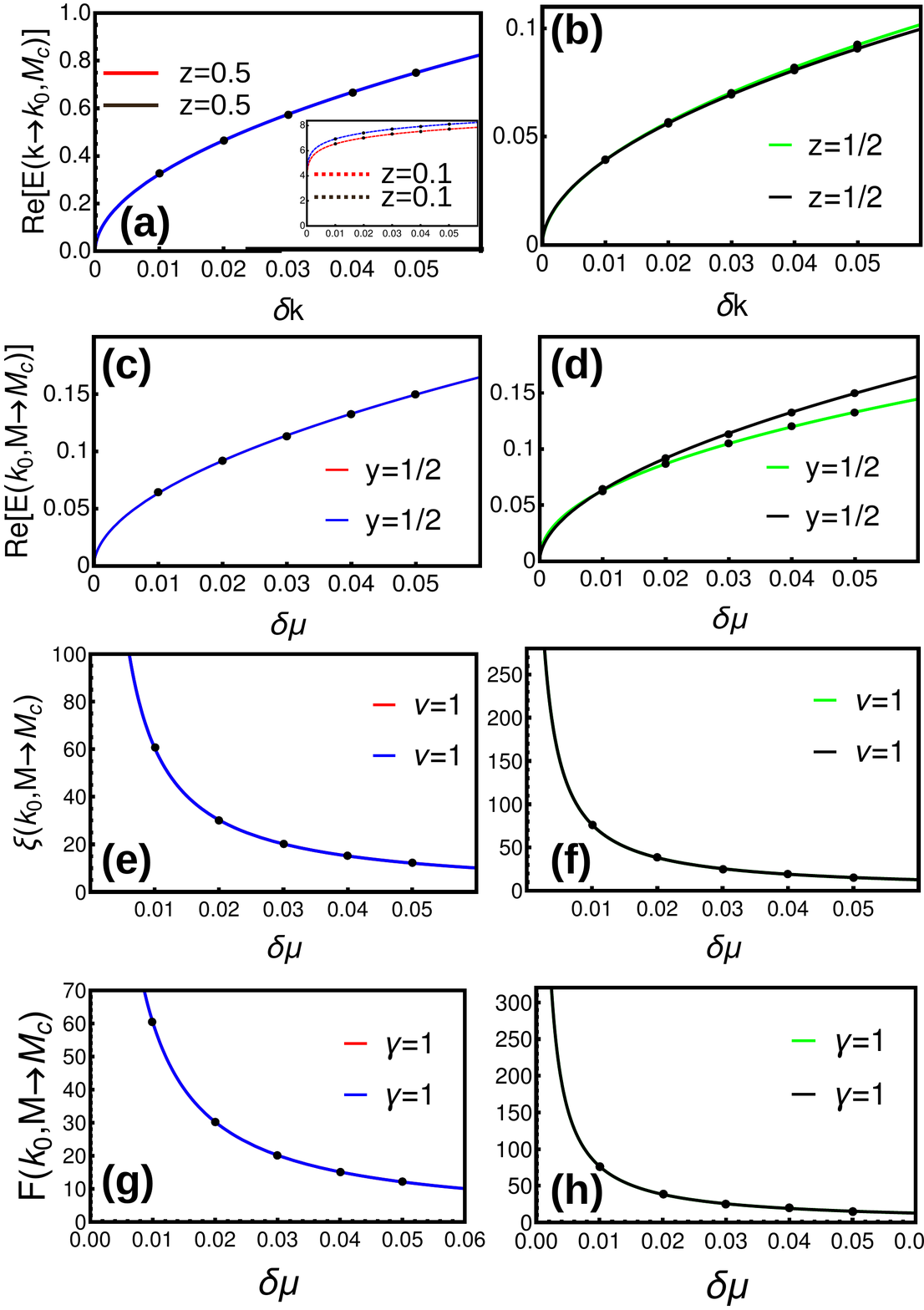}
	\caption{(Color online) Critical exponents through non-linear curve fitting method. The left and right panels represent the $k=0$ and $k=\pi$ criticalities respectively. The parameter space $\delta=0.2,t^{\prime}=1$ is considered. The red, blue, green and black colors represent the first ($\alpha=2.5,t_c=-1.14$), second (2.5,-1.54), third (2.5,1.06) and fourth (2.5,0.66) critical lines respectively. (a-b) Dynamical critical exponent. The inset represent the region $1<\alpha<2$ with parameter space ($\alpha=1.1,t_c=-10.38$) and (1.1,-10.78), where $z<0.5$. (c-d) Crossover critical exponent. (e-f) Localization critical exponent. (g-h) Susceptibility critical exponent.}
	\label{CritLong}
\end{figure}
Here we analyze the UCCE through nonlinear curve fitting method (Fig.~\ref{CritLong}). Around $k=0$, we obtain $z<0.5$ for the region $1<\alpha<2$, whereas $z=0.5$ for $\alpha>2$. In the region $1<\alpha<2$, we obtain undefined $\nu,\gamma$ and $y$ critical exponents, and $\nu=1,\gamma=1,y=0.5$ for $\alpha>2$. On the other hand, we get $z=0.5,y=0.5,\nu=1$ and $\gamma=1$ for all values of $\alpha$ around $k=\pi$. Table~\ref{ucll} summaries the critical exponents for different regions of non-Hermitian long-range SSH chain.
\subsection{Higher order bulk transitions:} To understand the order of phase transition, we check the derivatives of grand potential density for different orders. The phase transitions act as singularities in the density parameter space and the derivatives of density signals the order of transition with non-analytic curves. The polylogarithmic nature creates a singularity in the vicinity of $k\rightarrow0$, while the entire BZ contributes to the geometric phase. i.e., if $\omega_{singular}$ represents the grand potential density of the bulk, then it can be expressed as\cite{cats2018staircase} 
\begin{equation}
\omega_{singular}=\omega_x+\omega_y,
\end{equation}
where $\omega_x$ integral around $k=0$ and $\omega_y$ is the integral over the rest of the BZ respectively.
\begin{figure}[H]
	\centering
	\includegraphics[width=\columnwidth,height=6cm]{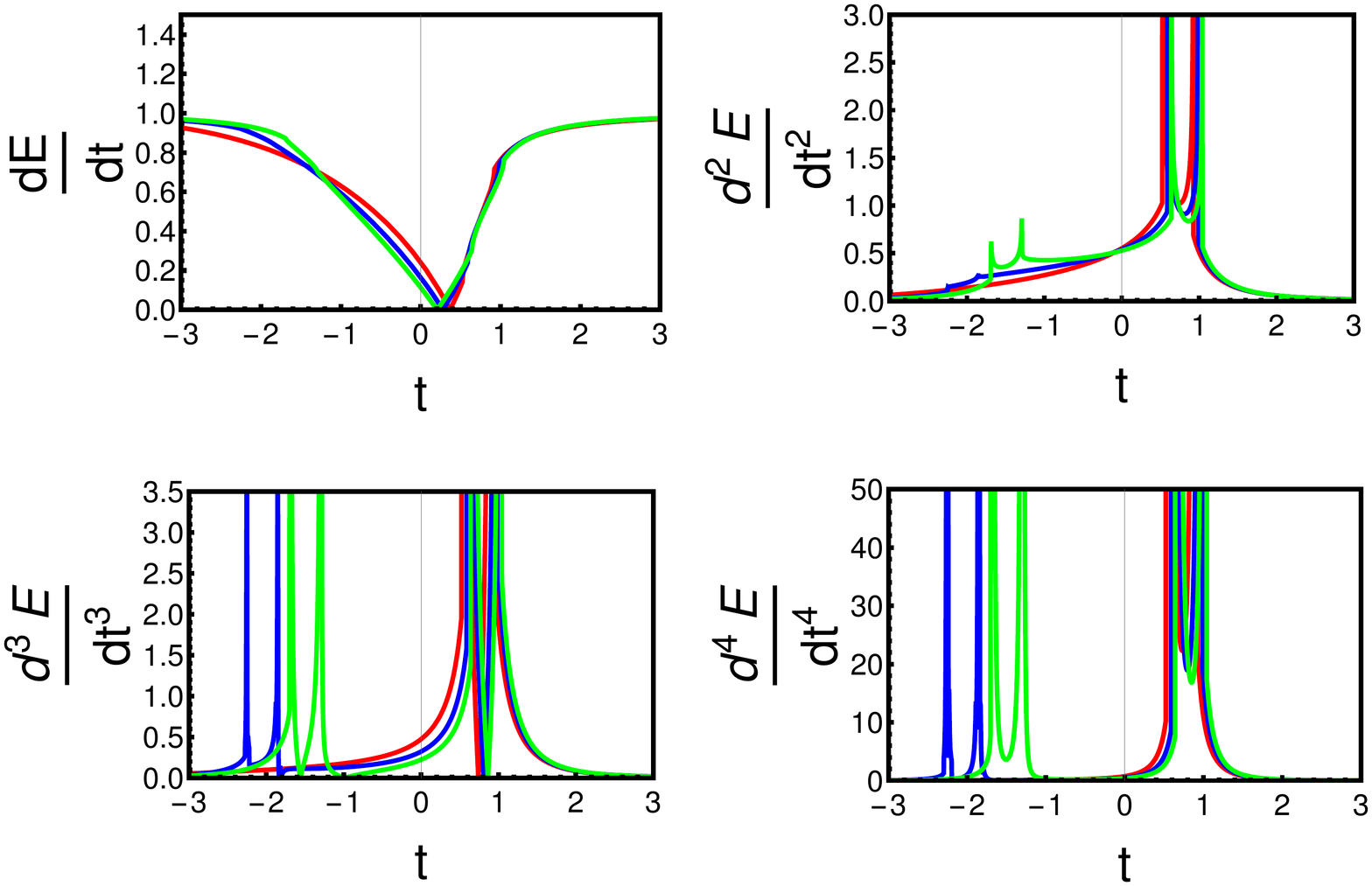}
	\caption{(Color online) Different order derivatives of ground state energy density. The singularities in the ground state represents distinction between the topological phases. The order of the transition represents the order of transition.}
	\label{OPT}
\end{figure}
 Through Eq.~\ref{ref}, we write an expanded term around $k=0$ as
\begin{equation}
\omega_x(k=0,\mathbf{M}(\alpha>2))=\int_{0-\epsilon}^{0+\epsilon}\sqrt{\mathbf{M}^2+k^2}dk\propto\mathbf{M}^2\ln|\mathbf{M}|.
\end{equation}
In this case we can observe a divergence in the second order derivative of the above equation, signaling a second order bulk transition. For the region $1<\alpha<2$,
\begin{equation}
\omega_x(k=0,\mathbf{M}(1<\alpha<2))\approx\int_{0-\epsilon}^{0+\epsilon}\sqrt{\mathbf{M}^2+k^{2(\alpha-1)}}dk.
\end{equation}
Through the Eq.~\ref{ref}, we observe the above term is proportional to $\Gamma(-\frac{\alpha}{2(\alpha-1)})\Gamma(\frac{2\alpha-1}{2(\alpha-1)})\mathbf{M}^{\frac{\alpha}{\alpha-1}}$, where $\Gamma$ function diverges for $\alpha=\frac{n}{n-1}$ where $n\in N$. Thus as one approaches $\alpha=1$ from the side of $\alpha=2$, the integral shows divergence for higher orders signaling a higher order bulk transitions.  Thus as we take the higher order derivatives, the criticalities corresponding to $k=\pi$ show the peaks while $k=0$ fails to produce the same. This signals a possible higher order phase transitions as one approaches $\alpha=1$. The region with integral WN (Fig.~\ref{long}) has similar Hermitian counterpart as explained in Ref.~\onlinecite{}, while the fractional WN do not have Hermitian counterpart. Interestingly, the fractional WN corresponding to $k=0$ shows a very narrow phase boundary, and as $\alpha\rightarrow1$, the non-Hermitian phase gradually vanishes. Thus at $\alpha=1$, the Hermitian phase exhibits an infinite order of bulk transition, while non-Hermitian phases shows almost none.
 This is also in good agreement with the hyper-scaling relation (Eq.~\ref{hyp}) also.
\begin{eqnarray}
	2-\alpha^*&=&1(1/2+1)=3/2\hspace{0.5cm}\text{for $k=\pi$},\nonumber\\
	2-\alpha^*&=&1(1/2+1)=3/2\hspace{0.5cm}\text{for $k=0,\alpha>2$},\nonumber\\
	2-\alpha^*&=&\text{Higher as $\alpha\rightarrow1$}\hspace{0.5cm}\text{for $k=0,1<\alpha<2$}.\nonumber
\end{eqnarray}
\begin{table}[H]
	\begin{center}

		\begin{tabular}{ |c|c|c|c|c|c|c|} 
			\hline
			$\hspace{0.2cm}\alpha\hspace{0.2cm}$ &$\hspace{0.2cm}k\hspace{0.2cm}$&$\hspace{0.2cm}z\hspace{0.2cm}$&$\hspace{0.2cm}\nu\hspace{0.2cm}$&  $\hspace{0.2cm}\gamma_{1,2}\hspace{0.2cm}$&$\hspace{0.2cm} y\hspace{0.2cm}$&
			$2-\alpha^*$\\ 
			\hline
			\hline
			
			$\alpha<1$	&0& - & -& -&-&-\\
			$1<\alpha<2$&& $z<0.5$ & -&-& 0.5&higher\\
			$\alpha>2$&& 0.5 & 1&1& 0.5&3/2\\
			\hline

			$\forall$$\alpha$	&$\pi$& 0.5 & 1&1& 0.5&3/2\\
			\hline
			\hline
		\end{tabular}
		
	\end{center}
	\caption{A comparison of universality class of critical 
		exponents of long-range Kitaev chain for different parameter spaces.}
	\label{ucll}
\end{table}
\section{Outlook and experimental aspects}\label{sec4}
Non-Hermitian topological systems are sensitive to boundary conditions, where the phase diagram may vary based on the choice of the boundary conditions. Here, we have worked on the periodic boundary condition, while the appearance of multi-criticality may be different in other conditions. We have observed the signature of localized modes at the gapless phases, which is an interesting phenomenon and merits further analysis in that regard.\\
Hermitian long-range models are the platform to realize the massive edge modes, which also can act as an effective topological qubit. Such possibility to explore massive edge modes in non-Hermitian counterparts can be interesting both from the perspective of theory and experiment. The field theoretical analysis of the Hermitian long-rang models reveals the breakdown of Lorentz invariance for the region $1<\alpha<2$. Our current work also signals such behavior through dynamical critical exponent, however, a detailed field theoretical analysis may answer this. \\
Long-range models are the field of interest from the experimental aspects also. The nature of shrinking characteristic length with the longer-range interaction helps to suppress the 
finite-size effects even by using a relatively small 
number of ions\cite{gong2016topological}. Long-range models have been realized in trapped ions~\cite{deng2005effective,britton2012engineered,hauke2010complete,roy2019quantum},
atom coupled to multi-mode cavities~\cite{douglas2015quantum}, magnetic impurities~\cite{zhang2019majorana,menard2015long} and quantum 
computation~\cite{amin2019information}. The effects of non-Hermiticity may give a different dimension towards the understanding of long-range systems.
\section{Discussion and Conclusion}\label{sec5}
Understanding the behavior of topological system with different range of coupling has attracted the attention of scientific community. The UCCE is an efficient tool to characterize the criticalities based on their behavior. The uniqueness of the multi-criticalities can be observed through UCCE. Here CRG is an useful tool to obtain the fixed and critical line configurations. But the CRG limits its coverage only to certain parameter spaces like HSP and non-HSPs corresponding to lower WNs. The symmetry plays a major role in defining the topological invariant. The current methodology may vary for the systems which lack chiral symmetry.\\ 
To conclude, we have provided a detailed analysis of criticality in longer and long-range non-Hermitian SSH chain. We have adopted the approach of separating the real and imaginary parts of the complex angle and defined the topological invariant. With the construction of staircase of transitions, we have observed the transition among even-even and odd-odd winding numbers, which actually has the Hermitian counterparts. To analyze the critical exponents, we have considered the $r=2$ case and calculated the universality class of critical exponents along with order of transition. The multi-critical point have shown some interesting feature, where the universality class is different than the rest of the parameter space. We adopt the CRG method to non-Hermitian systems and observe the fixed point configuration which actually depicts the difference in the universality class.\\
The long-range models are well known in Hermitian systems, due to the emergence of exotic particles like massive edge modes and physics of short-range,long-range inter-phase. Here we analyze model from momentum space characterization, and observe the interplay of polylogarithmic and non-Hermitian effect in winding vectors. 
%As the consequence of this, there occurs $W=1/4$, which are non-Hermitian phases representing the massive edge modes. 
Here we preferred the universality class of critical exponents to find the short-range limit rather than the topological invariant. CRG analysis gives the understanding of fixed/critical point configurations in long-range models. We observe the in the $\alpha\rightarrow1$ limit, it is possible to observe the higher order topological transitions.\\
The studies of long-range effects in non-hermitian models are less in literature. We believe, our work can be useful in understanding the interplay of long-range and non-Hermitian effects in topological state of matter.\\\\
\textit{Acknowledgments:}
SS would like to acknowledge DST (CRG/2021/000996) 
for the support and RRI library for the books and 
journals. 
YRK is grateful to AMEF for providing a PhD fellowship. The authors would like to acknowledge B S Ramachandra, C Sivaram, Rahul S and Ranjith Kumar R for useful 
discussions.
\bibliography{Non_Hermitian}
\appendix
\section{Derivation of winding number for non-Hermitian system}\label{appa}
The non-Hermiticity of the Hamiltonian reflects in the eigenvectors of the Hamiltonian as\cite{yin2018geometrical}
\begin{equation}
	W=\left(\frac{1}{2\pi}\right)\int_{-\pi}^{\pi}\frac{\langle u_k^L|i\partial_k|u_k^R\rangle}{\langle u_k^L|u_k^R\rangle}dk,
\end{equation}
where \begin{equation}
	\langle u_k^L|=\frac{1}{\sqrt{2}}
	\left(	\begin{matrix}
		\frac{\chi_x+i\chi_y}{\sqrt{\chi_x^2+\chi_y^2}}\\
		-1 
	\end{matrix}
	\right), \hspace{0.5cm}
	|u_k^R\rangle=\frac{1}{\sqrt{2}}
	\left(	\begin{matrix}
		\frac{\chi_x-i\chi_y}{\sqrt{\chi_x^2+\chi_y^2}}\\
		-1 
	\end{matrix}
	\right).
\end{equation}
Due to the non-Hermiticity, the winding vectors become complex (at least one of them) and the criticality condition is given by $h_x^2+h_y^2=0$. In Hermitian case, criticality occurs through single gap closing Dirac cone whereas in non-Hermitian case it is through (at least two) exceptional points. Due to the complex nature, the position of the exceptional point is given by,
\begin{eqnarray}
	\chi_x^{re}(k)=-\chi_y^{im}(k) &\hspace{0.5cm}\text{and}\hspace{0.5cm}& \chi_y^{re}(k)=\chi_x^{im}(k),\nonumber\\
	&\text{or}&\nonumber\\
	\chi_x^{re}(k)=\chi_y^{im}(k) &\hspace{0.5cm}\text{and}\hspace{0.5cm}& \chi_y^{re}(k)=-\chi_x^{im}(k).
\end{eqnarray}
In this model, we observe only two exceptional points. Thus we define a complex topological angle as,
\begin{equation}
	\phi=\phi^{re}+i\phi^{im}.
\end{equation}
Here the real and imaginary parts contribute to the argument and amplitude respectively. Thus the winding number is not affected by the imaginary part. Thus
\begin{eqnarray}
	e^{i2\phi}&=&\frac{1+i\tan(\phi)}{1-i\tan(\phi)}=\frac{\chi_x+i\chi_y}{\chi_x-i\chi_y},\nonumber\\
	e^{-i2\phi}&=&\left|\frac{\chi_x+i\chi_y}{\chi_x-i\chi_y} \right| ,\hspace{0.5cm}
	e^{-i2\phi_r}=\frac{\frac{\chi_x+i\chi_y}{\chi_x-i\chi_y} }{\left|\frac{\chi_x+i\chi_y}{\chi_x-i\chi_y} \right|},\nonumber\\
	\tan(2\phi_r)&=&\frac{Im(\frac{\chi_x+i\chi_y}{\chi_x-i\chi_y}) }{Re(\frac{\chi_x+i\chi_y}{\chi_x-i\chi_y} )}.
\end{eqnarray} 
which can be represented in a single equation as,
\begin{equation}
	\tan(2\phi_r)=\frac{\tan(\phi_1)+\tan(\phi_2)}{1-\tan(\phi_1)\tan(\phi_2)}=\tan(\phi_1+\phi_2),
\end{equation}
where $\tan(\phi_1)=\frac{\chi_y^{re}(k)+\chi_x^{im}(k)}{\chi_x^{re}(k)+\chi_y^{im}(k)}$ and $\tan(\phi_2)=\frac{\chi_y^{re}(k)-\chi_x^{im}(k)}{\chi_x^{re}(k)+\chi_y^{im}(k)}$. The angles $\phi_1$ and $\phi_2$ are real and they produce the winding number as
\begin{equation}
	W=\frac{1}{2}\left( \frac{1}{2\pi}\right) (\oint\partial_k\phi_1dk+\oint\partial_k\phi_2dk).\label{}
\end{equation}
For our model only $\chi_y$ contains the imaginary part. Thus the exceptional points are located at $(0, \chi_y^{im})$ and $(0, -\chi_y^{im})$. The wrapping of winding vectors $(\chi_x^{re}, \chi_y^{re})$ around first (second) exceptional point gives complex angle $\phi_1(\phi_2)$ with corresponding winding number $W_1(W_2)$ respectively. 
Due to the non-Hermiticity, each exceptional point induces its own origin of pseudo-spin space and corresponding winding vectors. Thus we work on extended winding vectors to understand the non-Hermitian effect in the parameter space. i.e.,
\begin{eqnarray}
	\chi_x(extended)=\chi_x^{re}(k)\pm\chi_y^{im}(k),\nonumber\\
	\chi_y(extended)=\chi_y^{re}(k)\pm\chi_x^{im}(k),\label{ext}
\end{eqnarray}
which correspond to parameter space $F_1(k,\mathbf{M})$ and $F_2(k,\mathbf{M})$ respectively.
\section{Detailed derivation of CRG equations}\label{appb}
Considering Eq.~\ref{rg}, we calculate the CRG flow equations to the longer-range model with $r=2$. 
Here $x1,x2$ and $x3,x4$ are critical and fixed lines respectively.The terms $X_{1,2},Y_{1,2}$ and $Z_{1,2}$ are given by
\begin{eqnarray}
	X_{1,2}&=&-\frac{2 \left(2^{1-\alpha }\mp1\right)^3}{\left(2^{-\alpha }-\delta +t\mp1\right)^3}-\frac{3 \left(\pm1-2^{2-\alpha }\right) \left(2^{1-\alpha }\mp1\right)}{\left(2^{-\alpha }-\delta +t\mp1\right)^2},\nonumber\\
	&+&\frac{\pm1-2^{3-\alpha }}{2^{-\alpha }-\delta +t\mp1}\nonumber\\
	Y_{1,2}&=&\frac{2^{1-\alpha }\mp1}{\left(2^{-\alpha }-\delta +t\mp1\right)^2},\nonumber\\
	Z_{1,2}&=&\frac{2^{-\alpha } \left(2^{1-\alpha }\mp1\right) \log (2)}{\left(2^{-\alpha }-\delta +t\mp1\right)^2}-\frac{2^{1-\alpha } \log (2)}{2^{-\alpha }-\delta +t\mp1}.\nonumber
\end{eqnarray}
%\begin{widetext}
1) In the vicinity of first exceptional point (parameter space corresponding to $F1(k,\mathbf{M})$), The CRG equations corresponding to $k=0$ are given by,
\begin{eqnarray}
	\frac{dt}{dl}&=&\left( \frac{1}{2}\right) \left(-\frac{X_2}{Y_2}\right) ,
	\frac{d\alpha}{dl}=\left( \frac{1}{2}\right) \left(\frac{X_2}{Z_2}\right).
\end{eqnarray}
with $x1=\frac{1}{2} (-2 \delta -1)\text{(red)},
x2=-\frac{1}{2^{\alpha }}-\delta -1\text{(blue)},
x3=2^{-\alpha }-\delta \text{(green)},
x4=-\frac{2^{\alpha } \delta -2^{\alpha }+8 \delta +1}{2^{\alpha }+8}\text{(black)}$.
\\\\
2) In the vicinity of first exceptional point (parameter space corresponding to $F1(k,\mathbf{M})$), The CRG equations corresponding to $k=\pi$ are given by,
\begin{eqnarray}
	\frac{dt}{dl}&=&\left( \frac{1}{2}\right) \left(-\frac{X_1}{Y_1}\right), 
	\frac{d\alpha}{dl}=\left( \frac{1}{2}\right) \left(\frac{X_1}{Z_1}\right),
\end{eqnarray}
with $x1=\frac{1}{2} (1-2 \delta )\text{(red)},
x2=-\frac{1}{2^{\alpha }}-\delta +1\text{(blue)},
x3=\frac{1}{2} \left(2^{1-\alpha }-\delta\right)\text{(green)},
x4=-\frac{2^{\alpha } \delta +2^{\alpha }-8 \delta +1}{2^{\alpha }-8}\text{(black)}.$\\\\
3) In the vicinity of second exceptional point (parameter space corresponding to $F2(k,\mathbf{M})$), The CRG equations corresponding to $k=0$ are given by,
\begin{eqnarray}
	\frac{dt}{dl}&=&\left( \frac{1}{2}\right) \left(-\frac{X_2}{Y_2}\right) ,
	\frac{d\alpha}{dl}=\left( \frac{1}{2}\right) \left(\frac{X_2}{Z_2}\right).
\end{eqnarray}
Here $x1=\frac{1}{2} (2 \delta -1)\text{(red)},
x2=-\frac{1}{2^{\alpha }}+\delta -1\text{(blue)},
x3=2^{-\alpha }+\delta\text{(green)},
x4=\frac{2^{\alpha } \delta +2^{\alpha }+8 \delta -1}{2^{\alpha }+8}\text{(black)}$.\\\\
4) In the vicinity of second exceptional point (parameter space corresponding to $F2(k,\mathbf{M})$), The CRG equations corresponding to $k=\pi$ are given by,
\begin{eqnarray}
	\frac{dt}{dl}&=&\left( \frac{1}{2}\right) \left(-\frac{X_1}{Y_1}\right), 
	\frac{d\alpha}{dl}=\left( \frac{1}{2}\right) \left(\frac{X_1}{Z_1}\right).
\end{eqnarray}
Here $x1=\frac{1}{2} (2 \delta +1)\text{(red)},
x2=-\frac{1}{2^{\alpha }}-\delta +1\text{(blue)},
x3=2^{-\alpha }+\delta \text{(green)},
x4=\frac{2^{\alpha } \delta -2^{\alpha }-8 \delta -1}{2^{\alpha }-8}\text{(black)}$.

\end{document}